\documentclass[12pt,preprint]{aastex}
\usepackage{natbib,psfig}
\def\lap{\lower.5ex\hbox{$\; \buildrel < \over \sim \;$}}
\def\gap{\lower.5ex\hbox{$\; \buildrel > \over \sim \;$}}
\begin{document}
\title{The Recent Star Formation History of the M31 Disk}
\author{ Benjamin F. Williams}
\affil{Harvard-Smithsonian Center for Astrophysics}
\affil{MS 67, 60 Garden Street, Cambridge, MA  02138}
\affil{williams@head-cfa.harvard.edu}

\begin{abstract}

The star formation history of the northern and southern M31 disk is
measured using samples of $BV$ photometry for 4' $\times$ 4' regions
taken from the KPNO/CTIO Local Group Survey \citep{massey2001}.  The
distances, mean reddening values, and age distributions of the stars
in these regions were measured using the routines of Dolphin (1997,
2002).  Independent measurements of overlapping fields show that the
results are stable for most samples.  A slight distance gradient is
seen across the major axis of the southern disk, and a mean distance
of 24.47$\pm$0.03 is found by combining the results.  Higher mean
reddening values follow the spiral structure.  The stellar
age distributions are consistent with episodic star formation confined
mainly to the gas-rich arm regions.  If these episodes were caused by
propagating density waves, the waves did not cause significant star
formation episodes in the gas-poor interarm regions.  Combination of
all of the results provides the total star formation rate for
1.4~deg$^2$ of the M31 disk for six epochs.  These results suggest
that star formation in the disk declined by $\sim$50\% from $\sim$250
to $\sim$50 Myr ago.  The lowest star formation rate occurred $\sim$25
Myr ago followed by a $\sim$20\% increase to the present.  The mean
star formation rate for this large portion of M31 over the past 60 Myr
is 0.63$\pm$0.07 M$_{\odot}$ yr$^{-1}$, suggesting a total mean rate
for the disk of $\sim$1 M$_{\odot}$ yr$^{-1}$.

%Extrapolation of this
%result to the entire disk area suggests a mean star formation rate of
%$\sim$1.1 M$_{\odot}$ yr$^{-1}$ for the entire M31 disk over the past
%100 Myr.
\end{abstract}

\keywords{galaxies: individual (M31); stellar content; evolution.}

\section{Introduction}

The stellar populations of a galaxy provide a fossil record of the
history of star formation activity within that galaxy.  In particular,
the distribution, colors and luminosities of the upper main sequence
and supergiant stars within the galaxy can provide a detailed history
of star formation within the galaxy over the past $\sim$200 Myr.  Once
this history is deciphered, it reveals the details of global star
formation processes within the galaxy, constraining its recent
evolution.  M31 provides a critical laboratory in which to study these
processes in a large spiral disk like the one in our Galaxy.
Knowledge of the star formation history (SFH) in the disk as a
function of location allows tests for propagation of star formation
across spiral arms, searches for coherent patterns of star formation
on large scales, and comparisons with structures seen at other
wavelengths, to look for correlations with star formation patterns.

The desire to understand how star formation has influenced the
evolution of M31 has led to studies too numerous to name
\cite[see][and references therein]{hodge1992}.  Studies date back to
\citet{hubble1929} who first examined the stars in M31.  On the other
hand, the ability to decipher the history of star formation among the
stellar complexes in the M31 disk has only recently been possible as
powerful data analysis tools have become available.  Ground-based
studies have been limited by photometric accuracy, field size, and
angular resolution.  Nevertheless, several studies
(e.g. \citealp{hodge1979,ivanov1985,magnier1992,magnier1997a,mochejska2000,massey2001,williams2003})
have put together large photometric samples of M31 stars and have
discovered age gradients in the disk
(e.g. \citealp{ivanov1985,magnier1997a}).  Resolved infrared stellar
photometry has suggested episodic star formation in some regions of
the disk \citep{kodaira1999}.

In recent years, Hubble Space Telescope (HST) observations have much
improved our knowledge of the stellar populations contained in the M31
disk.  The high angular resolution of HST has allowed deeper
photometry than was possible from the ground.  The deep,
high-resolution imaging has allowed age measurements of many young
star clusters in the disk
(e.g. \citealp{magnier1997b,williams2001a,williams2001b}).  At the
same time, studies of the field stars of the disk have put strong
constraints on the metallicity distribution and the SFH of many
locations within the disk
\citep{ferguson2001,sarajedini2001,williams2002,bellazzini2003}.

Unfortunately, the limited field of view of HST makes it difficult to
look for patterns of star formation in the M31 disk on large scales.
For this type of study we turn back to data sets from the ground,
which are now capable of covering large areas with a high degree of
homogeneity and photometric accuracy.  In this study I use the M31
data set from the CTIO/KPNO Local Group Survey \citep{massey2001} to
measure independent distances and mean extinctions in 345 regions of
the M31 disk, 106 of which overlap.  The SFH of each region is also
measured objectively and homogeneously, revealing patterns of star
formation in the M31 disk over the past 250 Myr.  The distance and
extinction measurements provide constraints for future observational
work in M31, and the star formation trends provide key insight about
the evolution of a spiral disk in the late stages of evolution.

\section{Data Acquisition and Reduction}

The data used for this project were generously supplied by the
CTIO/KPNO Local Group Survey (LGS) collaboration
(\citealp{massey2001};\\
\hbox{http://www.lowell.edu/users/massey/lgsurvey}) which is acquiring
1$''$ resolution, photometric data with the 8 k $\times$ 8 k Mosaic
cameras on the 4-m telescopes at KPNO and CTIO, entirely covering ten
LG galaxies in $UBVRI,$ and narrow-band H$\alpha$, [S~II]
($\lambda\lambda$6717, 6731) and [O~III] ($\lambda$5007).  The LGS is
working on their own, more rigorous, calibration of these data leading
toward a complete $UBVRI$ catalog of stars.  However, for the purposes
of this paper, I have simply used photometry from the literature to
perform a rough ($\sim$10\%) calibration.  The analysis used in this
paper is described in full detail, including tests of the photometry
routine, tests of the photometric calibration, and tests of the SFH
measurement technique, in \citet{williams2003} (hereafter W3).  An
incomplete brief summary of these details is supplied here for
convenience.

In short, the data consist of 6 fields from the MOSAIC camera on the
KPNO 4-meter telescope.  These fields cover most of the active
portions of the M31 disk.  The LGS team is using a total of 10 MOSAIC
fields to cover the entire M31 galaxy.  Here I analyze 6 of the 10,
covering a total area of 1.4 deg$^2$.  Much of the central 2 fields
will be dominated by the bulge.  These data therefore cover 6 of the 8
disk fields, or $\sim$75\% of the M31 disk.  The broadband images used
were in the Johnson $B$ and $V$ filters.  The dithered frames were stacked
and reduced using the DAOPHOT II and ALLSTAR packages \citep{stetson},
and the zero points for each field were determined using published
ground based photometry from previous surveys of \citet{mochejska2001}
and \citet{magnier1992}.

The full $V$, $B-V$ color-magnitude diagrams for the 6 fields analyzed in
this project are shown in Figure 1.  These diagrams show several
sequences typical of ground based photometry of the M31 disk performed
in previous surveys.  The upper main sequence of the M31 disk is seen
as a vertical sequence extending upward at $B-V\sim-0.1$.  Foreground
sequences extend upward from $B-V\sim0.5$ and $B-V\sim1.5$, showing
the Galactic halo and disk populations respectively.

I objectively dissected each field into subfields referred to as
``cells''.  Because F1 has fewer stars, being far out in the disk, it
was broken into 25 cells, 6.9$'$ $\times$ 6.9$'$ (1600 $\times$ 1600
pc) in size.  All other fields were broken into 64 cells, 4.3$'$
$\times$ 4.3$'$ (1000 $\times$ 1000 pc) in size.  Each of these cells
will be specified by their location on the grid by a field name
followed by pair of numbers x,y, where x is the number of grid cells
right of the left edge of the field and y is the number of grid cells
above the bottom of the field.  For example, cell F8 1,8 is the cell
in the far upper-left corner in field F8.

The resolved stellar photometry from each cell was used to measure the
distance, mean extinction and SFH of the cell.  In order to obtain a
precise understanding of the accuracy of the photometry as a function
of color and magnitude, a series of 150,000 artificial star tests were
performed in each region analyzed using the method described in W3.
In total, over 50 million artificial star measurements were performed
on the 6 fields analyzed in order to measure the completeness and
photometric accuracy as a function of color, magnitude and position in
the field.  The results of these tests were crucial for measuring
accurate SFHs with correct errors.

The outermost fields were used to obtain a sample of foreground stars
as described in W3.  This foreground sample was applied to each 4.3$'$
$\times$ 4.3$'$ sample in the M31 disk to statistically subtract
foreground contaminants as a function of color and magnitude.  These
foreground-subtracted $BV$ photometry samples of resolved stars in the
M31 disk were each run through the MATCH
\citep{dolphin1997,dolphin2002} routine for measuring the distance,
mean extinction and SFH of the region.  The software also measures the
metallicity distribution of the sample, but extensive tests, described
in W3, have shown that these measurements are not reliable for young
populations.

\section{Results}

\subsection{Distance and Extinction}

The SFH analysis routine provided distance and reddening information
for each cell.  The distances measured by the routine are plotted in
greyscale in Figure 2, where farther distances are darker gray.
Optical contours of M31 are overplotted for reference.  This plot
provides an opportunity to look for a distance gradient from the near
edge of the disk to the far edge.  If the active part of the disk is
$\sim$40 kpc across, with a mean distance modulus of 24.47 and an
inclination angle of 77 degrees the distance modulus of the near side
of the disk should be about 0.11 mag less than that of the far side.
Because the dust lanes on the western side of the disk appear to be in
the foreground of the bulge, this edge is currently believed to be the
closer one.

The distance parameter space explored by the measurements was limited
to between 24.4 and 24.5 magnitudes, which allowed more reliable SFHs
to be measured.  This limitation hinders the detection of a distance
gradient across the disk; however, such a detection is still possible
with precise relative photometry.  In the northern half of the disk,
the distance measurements are all the same within the errors,
revealing that the measurements do not have the precision necessary to
detect a gradient across the disk.  In the southern half of the disk,
it appears that the average distance measures to the southeast of the
major axis are systematically slightly higher than to the northwest of
the major axis.  The difference is not larger than the measured
errors, but the pattern suggests that the commonly accepted
orientation is correct: the southeastern disk is more distant.  When
all of these distance measurements are combined, a mean distance
measurement of 24.47$\pm$0.03 is obtained.
%The
%measured values in the far northwest and southeast corners of the
%fields should not be trusted because very few ($<100$) stars were
%involved in these measurements.

Along with the distance information, the routine outputs the most
likely mean extinction to the region, with errors.  In Figure 2 the
measured extinction values are plotted for 345 cells as a greyscale
map, with optical contours of M31 overplotted.  Measurements for
overlapping regions were averaged.  The measured values in the far
northwest and southeast corners of the fields should not be trusted
because very few ($<50$) stars were involved in these measurements.
Likewise, cells south of 40:15:00 all contained too few main sequence
stars to provide accurate reddening measurements.

An extinction gradient is seen from the inner disk to the outer disk,
decreasing outward.  In addition, high extinction appears to follow
the arm structure, which generally follows the outer contours of the
optical disk.  These areas also show strong dust emission
(e.g. \citealp{xu1996,haas1998}).  Because these areas also contain
most of the recent star formation (discussed in section 3.2), this
extinction pattern suggests that the production of dust in the high
density arm environment is more important than the destruction and
movement of dust by the UV fluxes and strong winds of massive young
stars.  As seen in most star forming galaxies and in star forming
regions of the Milky Way, dust and star formation generally occur
together.  While this result is not new, it provides important
confirmation that the analysis method is working properly.  The one
exception to this general rule is OB~78 (J2000 00:40:33.8, 40:44:22),
where it appears as though the slightly older age and intensity of the
star formation has blown away most of the surrounding gas and dust, as
seen in W3.

\subsection{Star Formation Histories}

The SFH of M31 as a function of position in the disk allows searches
for patterns of star formation in the disk on timescales comparable to
and shorter than an orbital period, before the structures are lost in
the complications of disk dynamics.  These measurements provide a
first look at the possible discoveries from such studies, although the
depth and accuracy limitations of ground-based photometry make strong
conclusions difficult to obtain.

Figures 3-8 show three-color images of the 6 fields in M31 analyzed.
Overplotted on these images are the SFHs output by MATCH for each cell
on the field.  The star formation rate for each epoch is shown with
red error bars within the cell.  The tickmarks on the abscissa and
ordinate provide the time period and star formation rate indicated by
the red error bars in each cell.  The interpretation of these SFHs for
each field is given below.

\subsubsection{F1}

Field F1, shown in Figure 3, was the outermost field analyzed, in
order to provide comparisons to the inner disk.  Since there are very
few young stars in these outer regions, a larger region size of
6.9'~$\times$~6.9' was used for each measurement.  The larger cell
sizes degraded the ability to look for star formation patterns, but
this sacrifice was necessary in order to obtain results with
statistical meaning.  

This field shows that most of the recent star formation is limited to
the inner disk.  A strong decrease in young stars is seen from OB~102
(J2000 00:46:35,41:10:00) outward.  A similar effect is seen in the
southern disk (F9), discussed below.  A rapid increase in star
formation is seen about 20 Myr ago in OB~102 (cell F1~4,2) along with
an overdensity of stars with ages $\lap$200 Myr, as seen in W3.  With
this larger picture, a decline in star formation is visible to the
southwest of the large OB association, revealing a possible
propagation of star formation from the southwest to the current
location of OB~102.

\subsubsection{F2}

Field F2 is shown in Figure 4.  Again the SFH around OB~102 is
measured; this time with smaller spatial bins.  The pattern to the
southwest of OB~102, detected in F1, is no longer seen.  Apparently,
large area sections were needed to detect this pattern; however, most
of the youngest stars appear to lie on the eastern side of the OB
association, consistent with the pattern seen in F1 (cells F2~3,6,
F2~4,6, F2~3,7).

This field also provides the first look at the SFH of the brighter
inner disk.  Significant activity is visible around OB~54 (J2000
00:44:30,41:51:00) in the northwest spiral arm and OB~49 (J2000
00:45:10,41:47:00) in the northeast spiral arm.  This enhanced star
formation appears to have been ongoing, though not constant, as far
back as the data are sensitive, suggesting that most of the star
formation in the disk has occurred in contained regions for several
hundred Myr.  This result was also found by \citet{bellazzini2003} and
\citet{williams2002}.  These age distributions could be caused by
propagating episodes of star formation, as discussed in section 4.3.

\subsubsection{F3}

Field F3, shown in Figure 5, contains several large groupings of hot,
blue stars.  The pattern seen among these most-recently active regions
of the disk appears to be one of prolonged star formation with a
recent burst beginning about 20-50 Myr ago.  This recent burst
explains the large populations of bright stars; however, the enhanced
star formation before the bursts, common in many of the arm regions,
is puzzling.  Interestingly, the SFHs in the arms tend to show a
decrease in star formation rate from $\sim$100 Myr to $\sim$50 Myr,
followed by an increase to the present.  The details of this pattern
are different for each region, but overall, it looks like these
regions are experiencing cycles of star formation on timescales of
10-200 Myr.  Such a pattern would be difficult to believe if seen in
only one or two arm regions, but when viewed in comparison with the
surrounding galaxy, conclusion are more easily drawn.
 
\subsubsection{F4}

Field F4, shown in Figure 6, was the innermost field analyzed.  This
field reaches into the bulge of M31, where the crowding becomes too
severe to perform reliable photometry from the ground.  This crowding
causes the large errors for star formation rates from 20-200 Myr ago
in the southwestern-most regions of this field.  Overall, the
significant recent star formation clearly follows the well-studied
spiral arms.

In conjunction with the results from field F3, an interesting pattern
emerges around OB~54 (J2000 00:44:30,41:51:00) which lies in cell
F3~5,4.  In F3 this OB association is well-centered in the cell and we
see the strong rise in star formation rate in this region over the
past $\sim$20 Myr, similar to other OB associations in the disk.  The
cell just north of the OB association appears to have recently
increased its star formation as well.  This increase is detected in
both the F3 and the F2 measurements.  At the same time, in F4, where
the southern half of the association is measured separately in cells
F4~2,7 and F4~3,7, it appears that star formation has recently
decreased in the southern half of this association, again suggesting
that the star formation is moving generally south to north, consistent
with what has been suggested in the southern disk and with the pattern
seen around OB~102 (see section 3.2.1).  This propagation is again
only seen over a small region, and may be revealing the process of
episodic star formation in these gas-rich regions.

\subsubsection{F8}

Field F8, shown in Figure 7, appears to contain most of the recent
star formation in M31.  This fact is accentuated by the dust spurs
that curl off to the west of the main disk.  Similar spurs have
recently been observed in more active spirals, one of the best
examples is M51 \citep{scoville2001}.  These spurs have been
well-modeled by dynamical simulations as occurring when differentially
rotating gas interacts with spiral arm structures \citep{kim2002}.
These structures are apparently signals of strong action caused by
spiral arms.

Studies of the Cepheids \citep{magnier1997a}, bright stars
\citep{ivanov1985} and star clusters \citep{williams2001a} in the
region around OB~78 suggest that star formation has been propagating
from south to north over the past $\sim$100 Myr.  No age gradient was
seen in the smaller sample measured in W3; however, this larger study
reveals a ramping down in star formation south of OB~78, a recent ramp
up at OB~78, and no strong trends to the north.  This pattern of star
formation is clearly a detection of the pattern seen in the star
clusters, bright stars, and Cepheids.  This strong wave of star
formation appears to be connected with the interaction of the gas disk
with the spiral arms, which also likely created the spur structures
seen to the west of the disk.

\subsubsection{F9}

Field F9, shown in Figure 8, contains very little activity in the
outer southwest region.  The decreasing number of young stars in the
southwestern cells confirms the radial age gradient in the disk.
Again the recent drop in star formation rate south of OB~78 is seen.
Without the results from F8, the drop would be less obvious since it
is only seen in the data point for the most recent epoch.
Fortunately, we can attach more significance to this data point
because of the results from the overlapping portion of field F8.

\section{Discussion}

In order to interpret the SFH measurement results, several comparisons
were performed.  Results for overlapping regions were quantitatively
compared to test for accuracy.  Results for the northern and southern
disk were compared to look for evolutionary differences across the
galaxy.  Results for the arm and interarm regions were compared to
look for the effects of propagating spiral density waves.  Finally,
all of the results were combined to look for large scale patterns in
star formation throughout the M31 disk.

\subsection{Comparison of Overlapping Frames}

%\subsubsection{Quantitative Comparison}

The independent measurements of similar areas in different fields
allowed important checks of the reliability of the measurement
technique.  There were 106 pairs of cells with centers within 3$'$ of
one another.  The consistency of the measurements was checked by
applying a $\chi^2$ analysis to the SFHs of each pair.  These $\chi^2$
measurements were accomplished by squaring the difference between the
measured star formation rates of each time bin.  The differences were
each divided by the sum of the measured square errors of each
measurement.  These measured square errors were tripled, as
experiments performed in W3 show that the errors tend to be
underestimated in these measurements.  The sum of these differences
was finally divided by the number of degrees of freedom (5 for these
SFHs) to give a reduced $\chi^2$ for each cell pair.  For convenience,
this reduced $\chi^2$ statistical comparison is simply called $\chi^2$
for the remainder of the paper.  The values obtained for all
overlapping fields are provided in Table 1, where column 1 provides
the cell pair, column 2 shows the angular distance between the cell
centers, column 3 gives the $\chi^2$ comparison, and column 4 provides
the $\chi^2$ value if one outlier is removed from the calculation.

The distribution of $\chi^2$ values for the overlapping cells was
compared to $\chi^2$ values measured for random cell pairs.  The two
distributions are shown in Figure 9.  The reduced $\chi^2$ values for
the overlapping cells clearly peak around a value of 1, while those of
random cell pairs are randomly distributed.  In total, 23\% of random
cell pairs had $\chi^2$ $<$3 while 64\% of overlapping cell pairs have
such values.  This result suggests that the details of the SFHs for
each cell are only reliable at about the 50\% level.  When combined
with results from nearby and overlapping cells, this confidence may be
improved, but conclusions should not be drawn based on the details of
the SFH of any one cell.  The 36\% of overlapping pairs that have high
$\chi^2$ values suggest that this issue should be investigated in more
detail, by comparing the general trends in SFH measured in overlapping
cells and investigating the typical causes for high $\chi^2$ values.

%\subsubsection{Qualitative Comparison}

Some overlapping cells with high $\chi^2$ values show agreement in
general SFH trend.  For example, 
%just north of OB~54, cell F2~8,3
%overlaps with cell F3~6,6 with $\chi^2$=2.65, even though both cells
%show a steady increase in star formation for this region over the past
%$\sim$100 Myr.  T
the rather constant rate in F8~7,2 is reproduced in F9~5,5 although
the $\chi^2$=6.51 because of discrepancies in early epochs.  The
increasing rate in F8~7,1 is reproduced in F9~5,4 ($\chi^2$ = 4.57),
and the high $\chi^2$ is again caused by discrepancies in early
epochs.  These early epochs are likely to be less reliable because
the measured rates depend heavily on the precise completeness near the
magnitude limit of the data.

There is generally good agreement between overlapping cells in regions
with large numbers of young stars.  For example, in the most active
area of F1, around OB 102, the closest overlapping cell pairs, F1~4,2
and F2~3,6 have a $\chi^2$ of 0.40.  To the southwest, cell pair
F1~5,1 and F2~5,4 have a $\chi^2$ of 0.70. All of the overlapping
cells west and south of OB~54 show nice agreement, as exemplified by
the striking similarity of the complex history shown for cell pair
F3~7,1 and F4~5,4 ($\chi^2$ = 1.55), showing a rapid increase in
activity over the past $\sim$20 Myr for this region. Comparison of the
cells F2~8,2 and F3~6,5, which contain the complex region of OB~54,
also obtain consistent results ($\chi^2$ = 1.02), showing enhanced
activity back to $\sim$100 Myr and further increased star formation
$\sim$10 Myr ago.  In addition, the cells south of OB~78 (F8~5,5 and
F9~3,8: $\chi^2$=1.63; F8~5,4 and F9~3,7: $\chi^2$=1.31) show
excellent overall agreement, revealing a mean decrease in star
formation in this region over the past 20 Myr.

Most poor $\chi^2$ values were due to the results of a single epoch.
If $\chi^2$ is calculated after removing a single outlier from each
SFH, 92\% of the overlapping pairs have $\chi^2 \leq 3$, and only 41\%
of random pairs match this well.  In order to assess whether these
single time-bin inconsistencies occurred because the boundaries of
overlapping cells were not perfectly aligned, we experimented with
aligning overlapping cells.  For example, cell F2~7,2 has a somewhat
different result than cell F3~5,5 ($\chi^2$ = 3.12).  Exactly matching
the cell regions improved the consistency of the measurements to
$\chi^2$ = 2.20; however, exclusion of the outlying data point at 100
Myr yields $\chi^2$ = 0.76. Apparently, the exact borders of the cells
do not strongly affect the consistency of results.  The
inconsistencies must arise from more subtle differences in the
photometry of the overlapping frames.  In any case, drastic changes in
star formation rate between adjacent epochs within a single cell are
rarely observed in overlapping frames.  Such changes are responsible
for most high $\chi^2$ values, and they should not be trusted unless
verified by a consistent measurement in adjacent or overlapping cells.
Conclusions cannot be drawn based on star formation rates measured
only once for single epochs.

Overall, the comparisons of overlapping cells reveal that SFHs
measured within individual cells are reliable at about the 50\% level.
The independent results for overlapping cells in active regions show
agreement ($\chi^2 < 2$), providing confidence that the measurement
technique is stable.  The confidence level increases for more recent
epochs, cells with greater numbers of bright stars, cells measured
independently in separate fields, and groups of cells showing similar
star formation patterns.  Most overlapping cell pairs with large
$\chi^2$ have one highly inconsistent data point.  Star formation
rates that deviate from general trends within a single SFH are
generally not reproduced in overlapping cells, and these deviant
points are typically responsible for high $\chi^2$ values in
overlapping cell pairs.  This problem shows that deviant rates
measured for individual epochs within single cells are not reliable.
 
\subsection{The Northern and Southern Regions}

Overall, the star formation of the northern and southern disks appears
to have been quite different over the past $\sim$200 Myr. Most
striking are the higher recent star formation rates measured for the
outer regions of the southern disk compared to those of the northern
disk.  This difference is seen most easily along the major axis, where
the star formation rates have been higher in the southwest corner of
F8 than in the northeast corner of F3 for the past 100 Myr.  The
difference is not entirely surprising considering the larger numbers
of open clusters observed in the southwestern half of the disk
\citep{hodge1979}; however, the result is surprising considering the
weaker gas emission from the southwestern disk \citep{devereux1994} and
the smaller number of OB associations in the southwestern half of the
disk \citep{vandenbergh1964a}.

In addition, the outer western half of the disk has been more active
recently than the outer eastern half.  The entire western side of
field F8 (cells F8~8,1 - F8~8,6) consistently shows significant star
formation over the past 100 Myr, while only 2 of the cells F3~1,2 -
F3~1,8 show significant activity over the same period.  A similar
difference is seen between the northwest disk and southeast disk,
where the northwest disk is more active.  This difference is unlikely
to be attributable to the fact that the western half is closer than
the eastern half because the young stars are bright and the extinction
is not strong in these outer regions.

Only 2 regions in the disk appear to have been forming stars at a high
rate for all 6 epochs.  Those areas are the area around OB~54 in the
northwest arm and the area around OB~78 in the southern disk.  The
star formation around OB~78 is certainly the stronger of the two.  It
is interesting to note that M32 and NGC~205, the nearest galaxies to
the M31 disk, lie to the south and northwest of the disk.  The
differences observed between the northern and southern disk may be due
to interaction with M32, which could be strong as seen in the tidal
stream recently detected around the southern part of the disk
\citep{ibata2001} and in models of the M31-M32 interaction
(e.g. \citealp{byrd1978}).  The higher rates in the northwest could
have been caused by interactions with NGC~205 and/or M32.  This
interpretation is certainly speculative, but the locations of these
more active outer disk regions are curiously consistent with such a
scenario.

\subsection{The Arm and Interarm Regions}

Our results reveal that the arm and interarm regions of M31 have been
quite distinctive for most of the past 250 Myr.  While the star
formation rate in the arms has been relatively episodic, changing
noticeably from epoch to epoch, the interarm regions have been
relatively inactive the entire time.  This greater amount of star
formation in the arms has likely been possible because of the greater
gas content of the arm regions.  These regions have been observed to
have enhanced gas content in many studies (e.g. \citealp{brinks1984}
(H I); \citealp{dame1993} (CO); and many others).

Only in epoch 4 were the star formation rates of the arm and interarm
regions similar.  This similarity occurred because the star formation
rate in the arms became quite low.  In fact, this epoch contained the
least amount of star formation of any epoch measured (see Table 2 and
Figure 10).  The long-lived inactivity in the interarm regions
suggests that, if density waves propagated through these regions, the
density enhancements were not strong enough to overcome the low
ambient gas density and significantly increase star formation.  The
propagation of star formation through some gas-rich regions is
suggested by the global patterns of star formation discussed in the
next section.  These results suggest that strong star formation
activity is not likely to propagate far through the disk because the
gas-poor interarm regions do not support significant star formation
episodes even under the influence of a density wave passage.

\subsection{Global Properties of the Recent SFH}

Perhaps the most useful application of these results is seen in Figure
10, where the results for the star formation rate during each epoch in
each cell have been combined into greyscale maps. In these maps,
darker gray areas have higher star formation rates.  Each epoch probed
by the analysis is shown separately.  Results for overlapping regions
of cells were averaged to make the best use of the multiple
measurements in those cells.  Each map is overplotted with an optical
contour map of M31 so that it is easy to see the location of the
galaxy in these star formation rate maps.  An overall assessment of
these images is described in Table 2, where the total star formation
rate calculated for these regions is shown for each epoch measured.  A
movie version of Figure 10 is available on the world wide web at
\hbox{http://www.astro.washington.edu/ben/m31\_sfh.gif}.

Globally, the images and the table both reveal a decrease in star
formation from $\sim$200 Myr ago to $\sim$50 Myr ago.  The difference
lies in the intensity of the star formation in the active regions.
Both the northern and southern spiral arms show more intense activity
in epochs 1 and 2.  Tests of the SFH measurement technique performed
and described in W3, suggest that the star formation rate in epoch 1
may be systematically over-estimated in these results because of the
dependence of this measurement on photometry near the magnitude limit
of the data.  However, the measurements for epoch 2 performed well in
the same tests.  In addition, the decrease in star formation rate does
not occur in all of the active regions, decreasing the likelihood that
it is the result of systematic errors. For example, the region south
of OB~78 (J2000 0:40:00, 41:30:00) does not show this drop.  The fact
that the strong decrease in the star formation rate between epochs 2
and 3 is likely to be real suggests that significant changes in the
global star formation rate can occur on timescales of $\sim$100 Myr.

Another obvious change seen during these epochs is a steady decrease
in star formation in the bulge region.  These measurements are not
reliable.  The rates measured for the bulge region are highly
uncertain in the older epochs due to the high crowding in this region
(see Figure 6).  These large errors allow for higher star formation
rates in the earlier epochs.

The details of these maps are best analyzed by blinking through the
epochs and watching for areas where the star formation shows evidence
for bursts or propagation.  This viewing option is available in a
movie located on the world wide web at\\
\hbox{http://www.astro.washington.edu/ben/m31\_sfh.gif}.  A
description of the differences as seen by such a technique will be
discussed.  Between epochs 1 and 2, activity in the northeast spiral
arm decreased while that in the southeast spiral arm increased.  Then,
between epochs 2 and 3, the activity of the northeast spiral arm
continued to decrease, and the activity in the southeast arm also
decreased.  Here, a passage of a spiral density wave is suggested in
epochs 1, 2 and 3, as star formation left the northeast spiral arm,
entered the southeast arm, and then exited the southeast arm.

During epochs 1-3, the region around OB~54 (J2000 00:42:50,41:37:00;
the northwest spiral arm) was undergoing a similar experience to the
southeast arm.  This area shows little activity in epoch 1, then it
shows a burst of activity covering epochs 2 and 3 before settling down
in epoch 4.  These findings suggest the passage of a density wave
through this region over a similar time period as one passing through
the southeast arm, indicating symmetry in these two areas.

In epochs 3 and 4, the disk as a whole went through its weakest period
of star formation.  While the arms continued to form stars at a lower
rate, most of the strong star formation became confined to a few large
OB associations (OB~48, OB~49, OB~54, OB~78, OB~102).  Then, in epoch
5, activity increased again in both the northeast and southeast arms.
The regions southeast and east of OB~78 also began to form more stars,
and OB~102 became strongly active.

Finally, from epoch 5 to epoch 6, there were several significant
changes in activity.  Epochs 5 and 6 were only 8 Myr long and 4 Myr
long respectively, meaning that these changes took place on a
timescale of $<$10 Myr. In this period of time, the star formation
rate in the northeast spiral arm increased.  This recent increase was
also seen in W3.  The strong star formation south of OB~78 which
pervaded all of the other epochs, finally fell off, likely due to gas
depletion.  The star formation in this region appears to have moved
northward in this epoch, remaining strong in OB~78 and just northward,
consistent with the pattern seen in Cepheid ages by
\citet{magnier1997a} and in star cluster ages by
\citet{williams2001a}.  Activity in OB~102 appears to have moved
slightly eastward.  The area of OB~59 became active again for the
first time since epoch 3.

All of these changes seen in the SFH of the M31 disk suggest that some
OB associations and spiral arms go through multiple episodes of star
formation, likely attributable to the passage of density waves through
these gas-rich regions of the disk.

Because the total star formation rate appears to have stayed fairly
stable over the past $\sim$64 Myr, the measurements for epochs 3-6
were averaged in order to measure one precise mean recent star
formation rate covering 1.4 deg$^2$ ($\sim$75\%) of the M31 disk.  The
resulting recent star formation rate is 0.63$\pm$0.07 M$_{\odot}\
yr^{-1}$.  Extrapolating this value to the whole disk suggests a total
rate of $\sim$1 M$_{\odot}\ yr^{-1}$, which is in good agreement with
the far infrared luminosity of M31 \citep{devereux1994}, which
provides a current star formation rate of $\lap$1 M$_{\odot}\ yr^{-1}$
using the conversion technique of \citet{inoue2000}.

\section{Conclusions}

The SFH of most of the M31 stellar disk has been measured back to 250
Myr ago using $BV$ photometry of 6 fields taken with the MOSAIC camera
on the 4-m telescope at KPNO.  The distance, mean extinction, and
stellar age distribution were determined independently for each 18.5
arcmin$^2$ portion of the fields (47.3 arcmin$^2$ portions in the
outermost field).  Reliability of these measurements was assessed
based on the consistency of measurements in overlapping areas of the 6
fields and on tests of the measurement technique described in an
earlier publication (W3).

The distances of all regions of the disk were consistent within the
measured errors; however, the distances measured in the southern half
of the disk provide a hint that the disk southeast of the major axis
is the more distant, in agreement with the currently accepted
orientation.  Combining all 345 distance results gives a distance
modulus to M31 of 24.47$\pm$0.03.  This result is biased by the input
distance modulus range of 24.4 to 24.5; however, these results
constrain the distance to the upper end of the input range.  This
distance is also identical to several other recent measurements
(e.g. \citealp{stanek1998}, and \citealp{holland1998}).

The mean extinction measurements show that the areas of higher
extinction tend to follow the spiral arms, with the exception of the
area around OB~78, where the dust appears to be depleted.  This
depletion was seen in W3, who suggested that the dust may have been
blown out by the strong winds of large numbers of hot young stars.

The southern disk appears to have been more active in recent epochs
than the northern disk, and the outer western disk appears to have
been more active in recent epochs than the outer eastern disk.  Areas
of the disk showing persistent, strong activity appear in areas
possibly influenced by interactions with NGC~205 and/or M32.

There is evidence of star formation propagation across some gas-rich
regions of the disk.  The strongest evidence favors a process of star
formation that has been episodic and confined mainly to gas-rich
regions of the arms for the past few hundred Myr.  If these episodes
have been caused by propagating density waves, the waves appear not to
have caused episodes of star formation as they passed through gas-poor
interarm regions.

Combination of all of the results provides a global SFH for most of
the M31 disk for the past 250 Myr.  The total star formation rate
appears to have decreased by about 50\% from $\sim$250 to $\sim$50 Myr
ago.  The lowest global rate then occurred about 25 Myr ago followed
by a $\sim$20\% increase over the past $\sim$20 Myr.  Overall, the
mean total star formation rate for this 1.4 deg$^2$ ($\sim$75\%) of
the M31 disk over the past 64 Myr has been 0.63$\pm$0.07 M$_{\odot}$
yr$^{-1}$.  Scaling this value to the rest of the disk provides an
estimated star formation rate of $\sim$1 M$_{\odot}$ yr$^{-1}$ for the
M31 disk.

I thank the LGS for supplying the data used for this project,
especially Phil Massey for his work in observing, preparing and
providing the images for the photometric measurements.  I thank Andrew
Dolphin for supplying the software package necessary for the SFH
analysis.  Support for this work was provided in part by NASA through
grant number GO-9087 from the Space Telescope Science Institute and by
the Chandra X-ray Center under grant GO2-3103X.

%\bibliography{apjmnemonic,references}
%\bibliographystyle{apj}

\begin{figure}
\centerline{\psfig{file=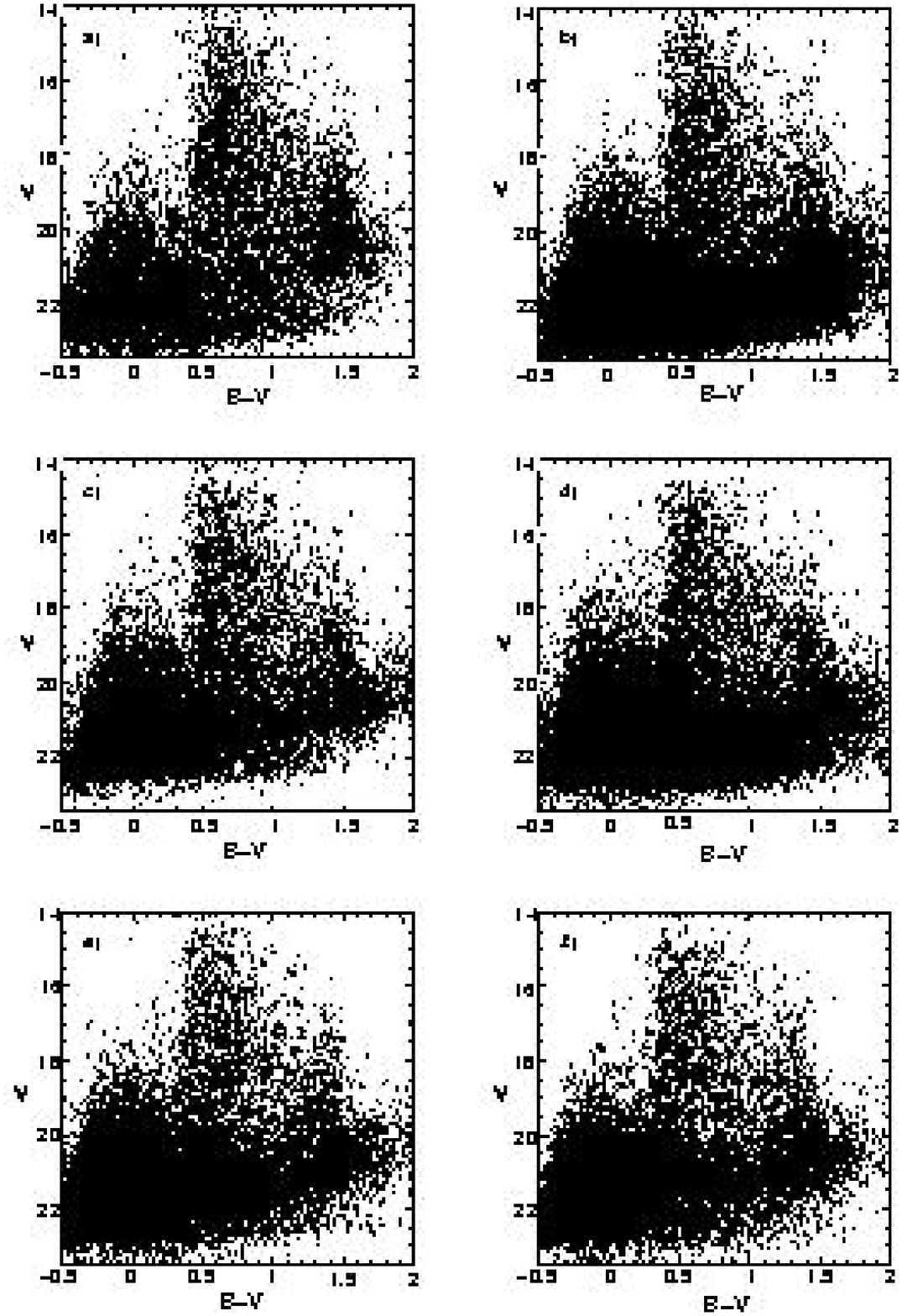,height=8.0in,angle=0}}
\caption{$V$, $B-V$ color-magnitude diagrams of all stars in all 6 LGS fields
analyzed: a) F1, b) F2, c) F3, d) F4, e) F8 and F) F9.}
\end{figure}

\begin{figure}
\centerline{\psfig{file=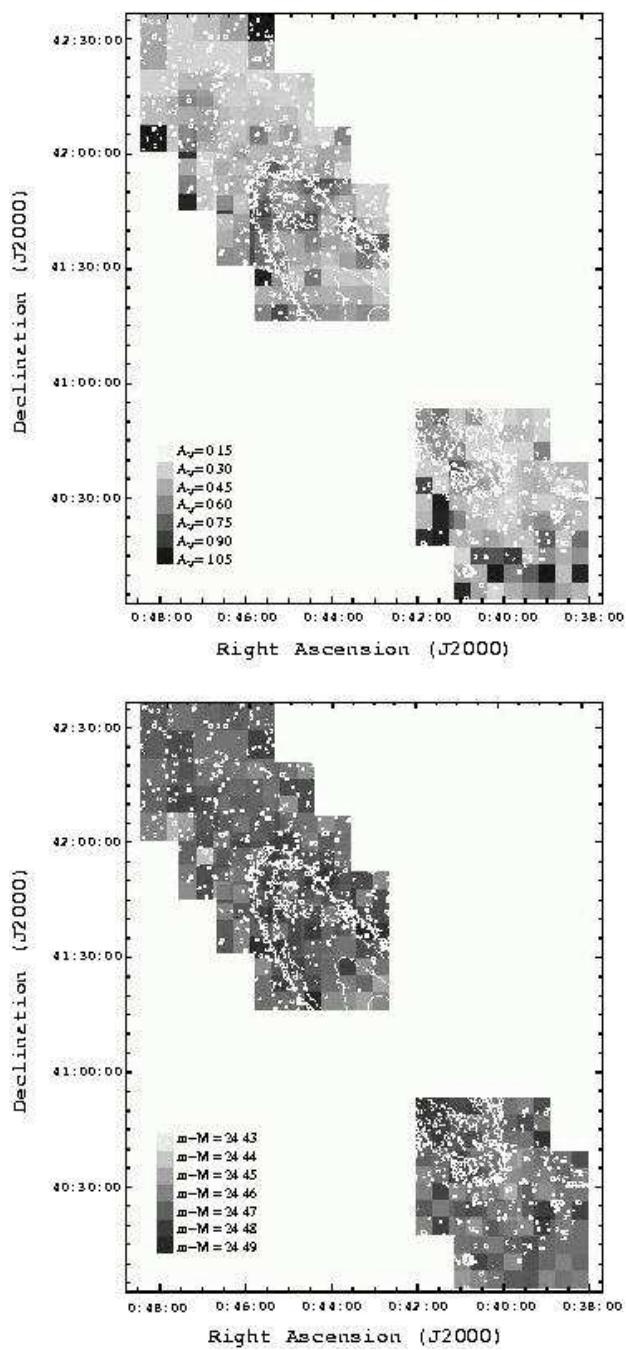,height=7.0in,angle=0}}
\caption{Top: Greyscale shows the measured extinction ($A_V$) in M31
as a function of position in the disk.  A gradient is seen across the
minor axis, most pronounced in the north.  The arm regions appear to
show higher extinction than interarm.  Contours show the location of
the brightest part of the disk.  Bottom: Greyscale shows the measured
distance moduli of the M31 disk as a function of position.  No
gradient is detected in the northern disk. In the southern disk
distances to the southeast of the major axis appear slightly (but not
significantly) greater than those to the northwest of the major axis.
Contours show the location of the brightest part of the disk.
Measured values in the far northwest and southeast corners of the
fields are highly uncertain due to a low number of young stars in
these areas.}
\end{figure}

\begin{figure}
\centerline{\psfig{file=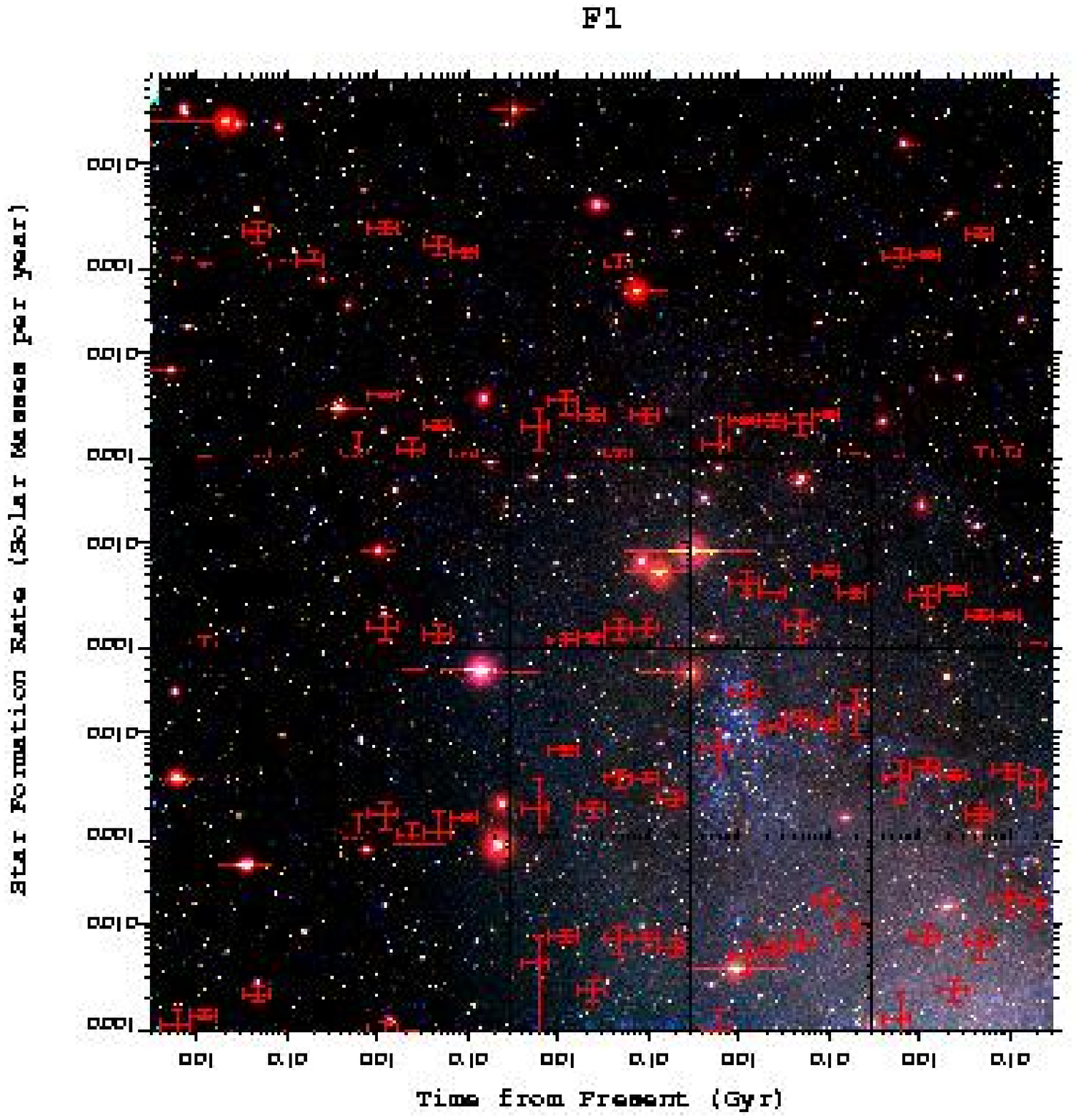,height=7.0in,angle=0}}
\caption{A color image of field F1 is divided into 25 6.9$'$ $\times$
  6.9$'$ cells.  Overplotted on each 6.9$'$ $\times$ 6.9$'$ cell
  is its measured SFH in M$_{\odot}\ yr^{-1}$ back to 256 Myr (see
  axes and tickmark labels).  Red error bars mark the star formation
  rates for each time period in each cell as measured by MATCH.  In
  each cell, the abscissa provides the time period explored, and the
  ordinate provides the star formation rate.  The image is a composite
  of the B band (blue), the V band (green) and the I band (red) images
  from the LGS data set (Massey et al. 2001).}
\end{figure}

\begin{figure}
\centerline{\psfig{file=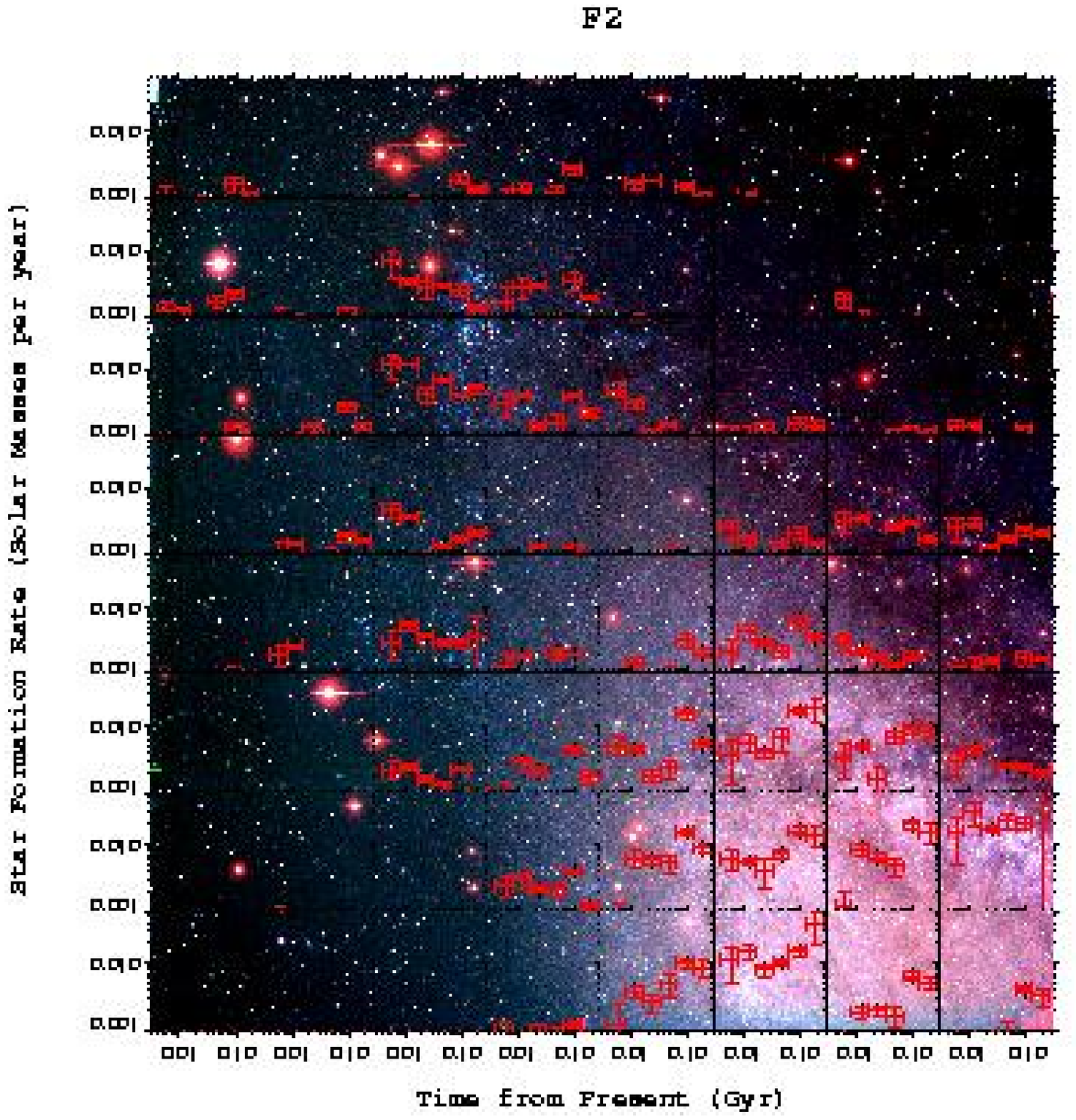,height=7.0in,angle=0}}
\caption{A color image of field F2 is divided into 64 4.3$'$ $\times$
  4.3$'$ cells.  Overplotted on each 4.3$'$ $\times$ 4.3$'$ cell
  is its measured SFH in M$_{\odot}\ yr^{-1}$ back to 256 Myr (see
  axes and tickmark labels).  Red error bars mark the star formation
  rates for each time period in each cell as measured by MATCH.  In
  each cell, the abscissa provides the time period explored, and the
  ordinate provides the star formation rate.  The image is a composite
  of the B band (blue), the V band (green) and the I band (red) images
  from the LGS data set (Massey et al. 2001).}
\end{figure}

\begin{figure}
\centerline{\psfig{file=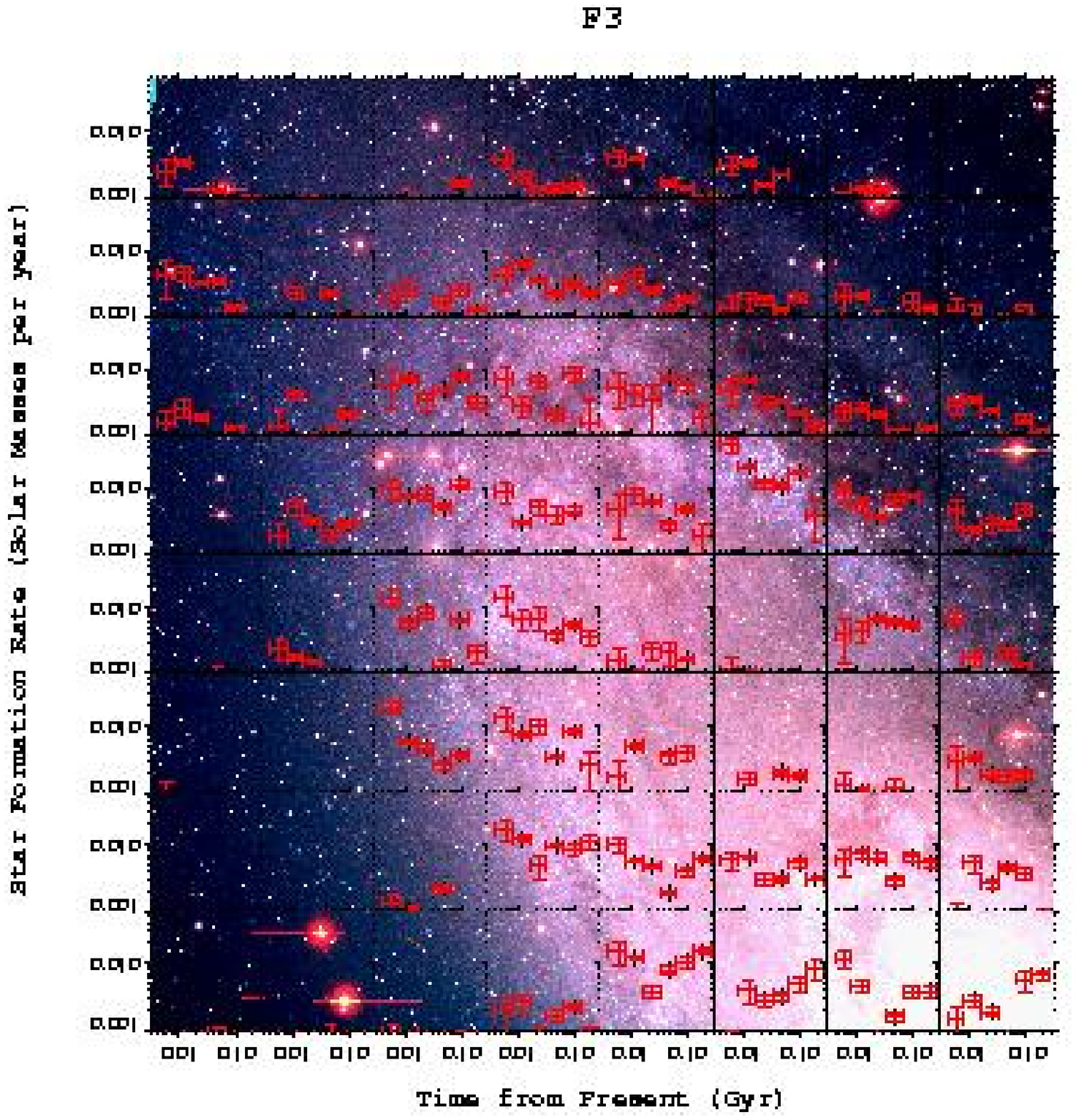,height=7.0in,angle=0}}
\caption{A color image of field F3 is divided into 64 4.3$'$ $\times$
  4.3$'$ cells.  Overplotted on each 4.3$'$ $\times$ 4.3$'$ cell
  is its measured SFH in M$_{\odot}\ yr^{-1}$ back to 256 Myr (see
  axes and tickmark labels).  Red error bars mark the star formation
  rates for each time period in each cell as measured by MATCH.  In
  each cell, the abscissa provides the time period explored, and the
  ordinate provides the star formation rate.  The image is a composite
  of the B band (blue), the V band (green) and the I band (red) images
  from the LGS data set (Massey et al. 2001).}
\end{figure}

\begin{figure}
\centerline{\psfig{file=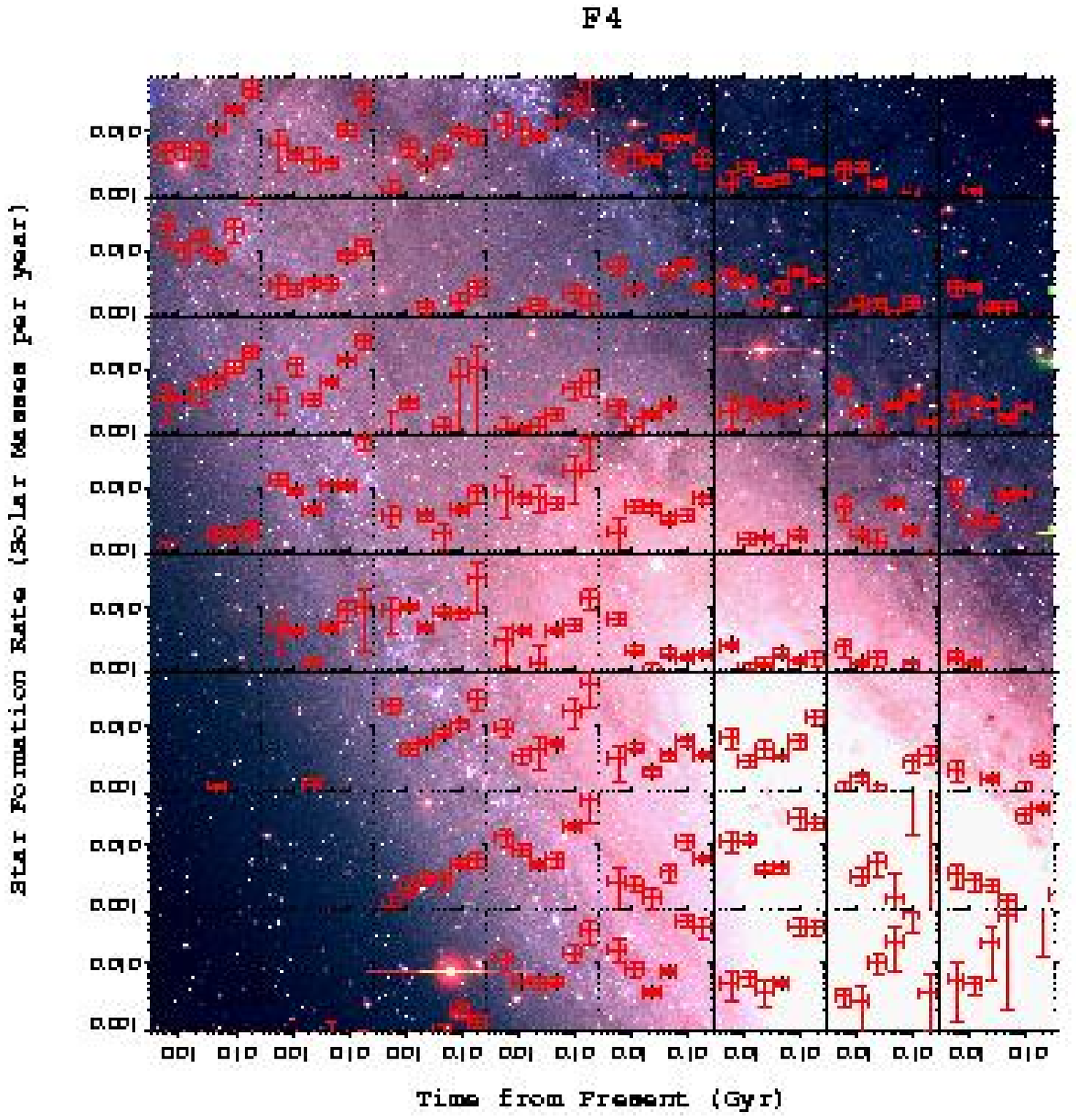,height=7.0in,angle=0}}
\caption{A color image of field F4 is divided into 64 4.3$'$ $\times$
  4.3$'$ cells.  Overplotted on each 4.3$'$ $\times$ 4.3$'$ cell
  is its measured SFH in M$_{\odot}\ yr^{-1}$ back to 256 Myr (see
  axes and tickmark labels).  Red error bars mark the star formation
  rates for each time period in each cell as measured by MATCH.  In
  each cell, the abscissa provides the time period explored, and the
  ordinate provides the star formation rate.  The image is a composite
  of the B band (blue), the V band (green) and the I band (red) images
  from the LGS data set (Massey et al. 2001).}
\end{figure}

\begin{figure}
\centerline{\psfig{file=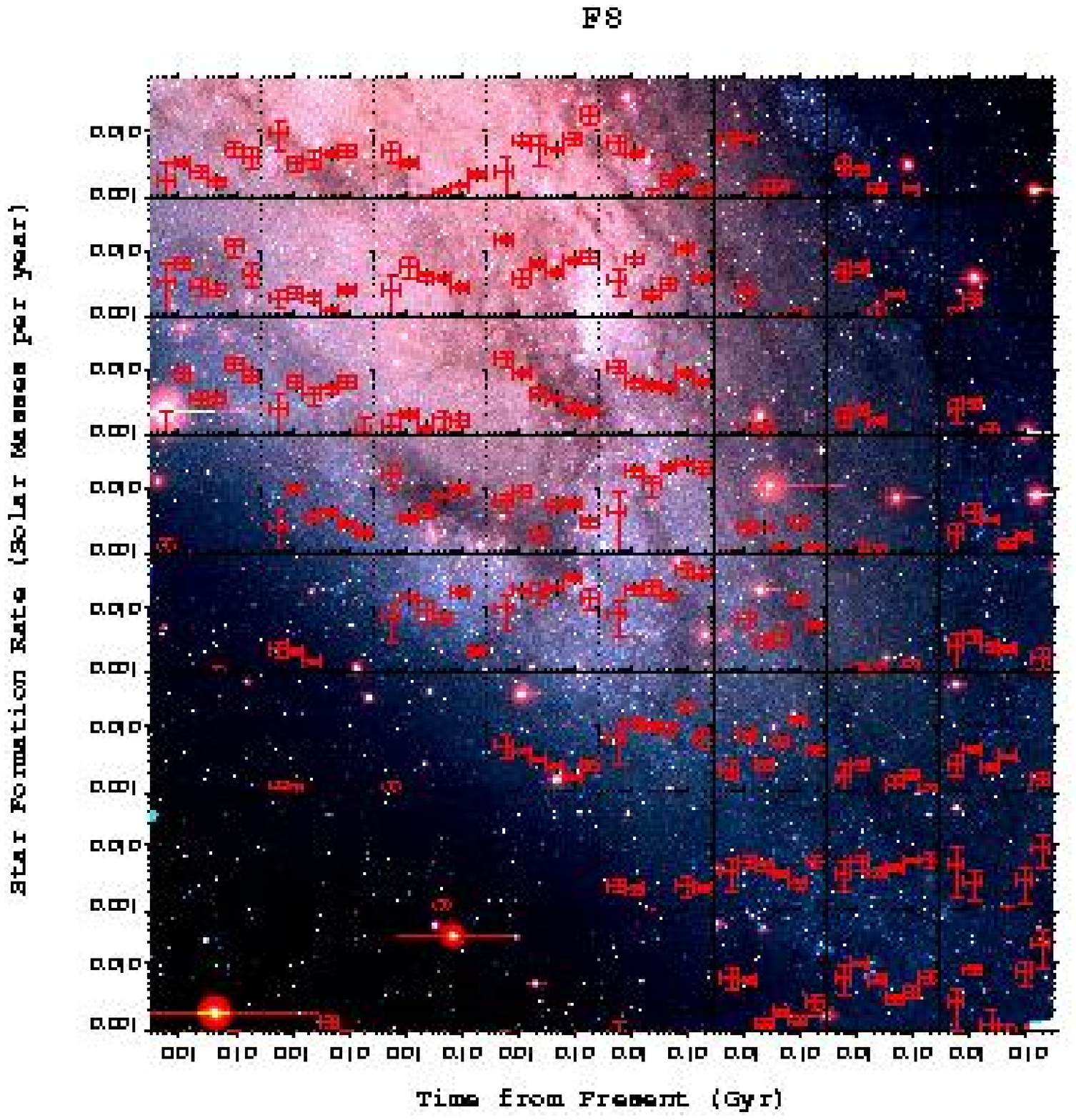,height=7.0in,angle=0}}
\caption{A color image of field F8 is divided into 64 4.3$'$ $\times$
  4.3$'$ cells.  Overplotted on each 4.3$'$ $\times$ 4.3$'$ cell
  is its measured SFH in M$_{\odot}\ yr^{-1}$ back to 256 Myr (see
  axes and tickmark labels).  Red error bars mark the star formation
  rates for each time period in each cell as measured by MATCH.  In
  each cell, the abscissa provides the time period explored, and the
  ordinate provides the star formation rate.  The image is a composite
  of the B band (blue), the V band (green) and the I band (red) images
  from the LGS data set (Massey et al. 2001).}
\end{figure}

\begin{figure}
\centerline{\psfig{file=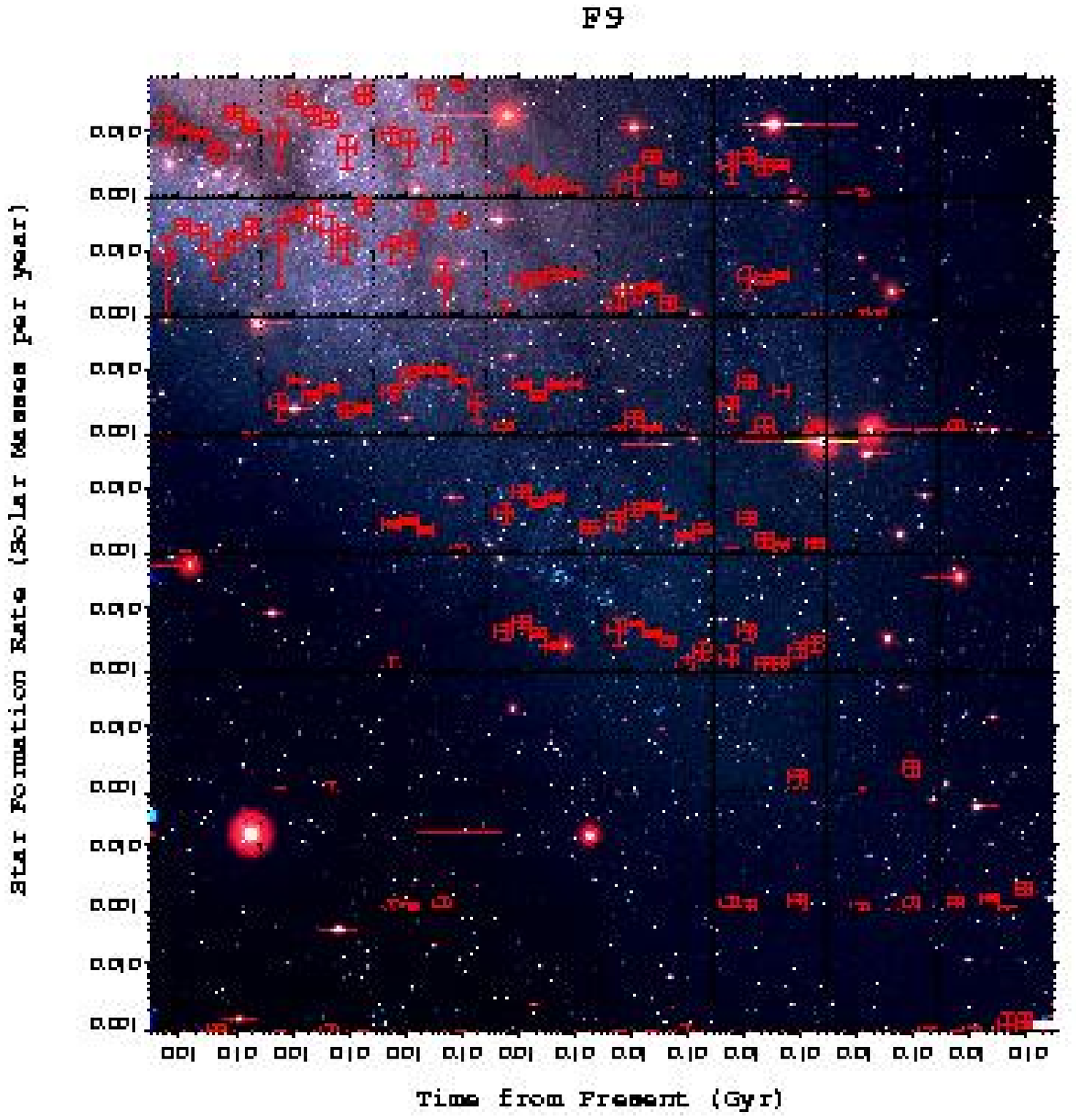,height=7.0in,angle=0}}
\caption{A color image of field F9 is divided into 64 4.3$'$ $\times$
  4.3$'$ cells.  Overplotted on each 4.3$'$ $\times$ 4.3$'$ cell
  is its measured SFH in M$_{\odot}\ yr^{-1}$ back to 256 Myr (see
  axes and tickmark labels).  Red error bars mark the star formation
  rates for each time period in each cell as measured by MATCH.  In
  each cell, the abscissa provides the time period explored, and the
  ordinate provides the star formation rate.  The image is a composite
  of the B band (blue), the V band (green) and the I band (red) images
  from the LGS data set (Massey et al. 2001).}
\end{figure}

\begin{figure}
\centerline{\psfig{file=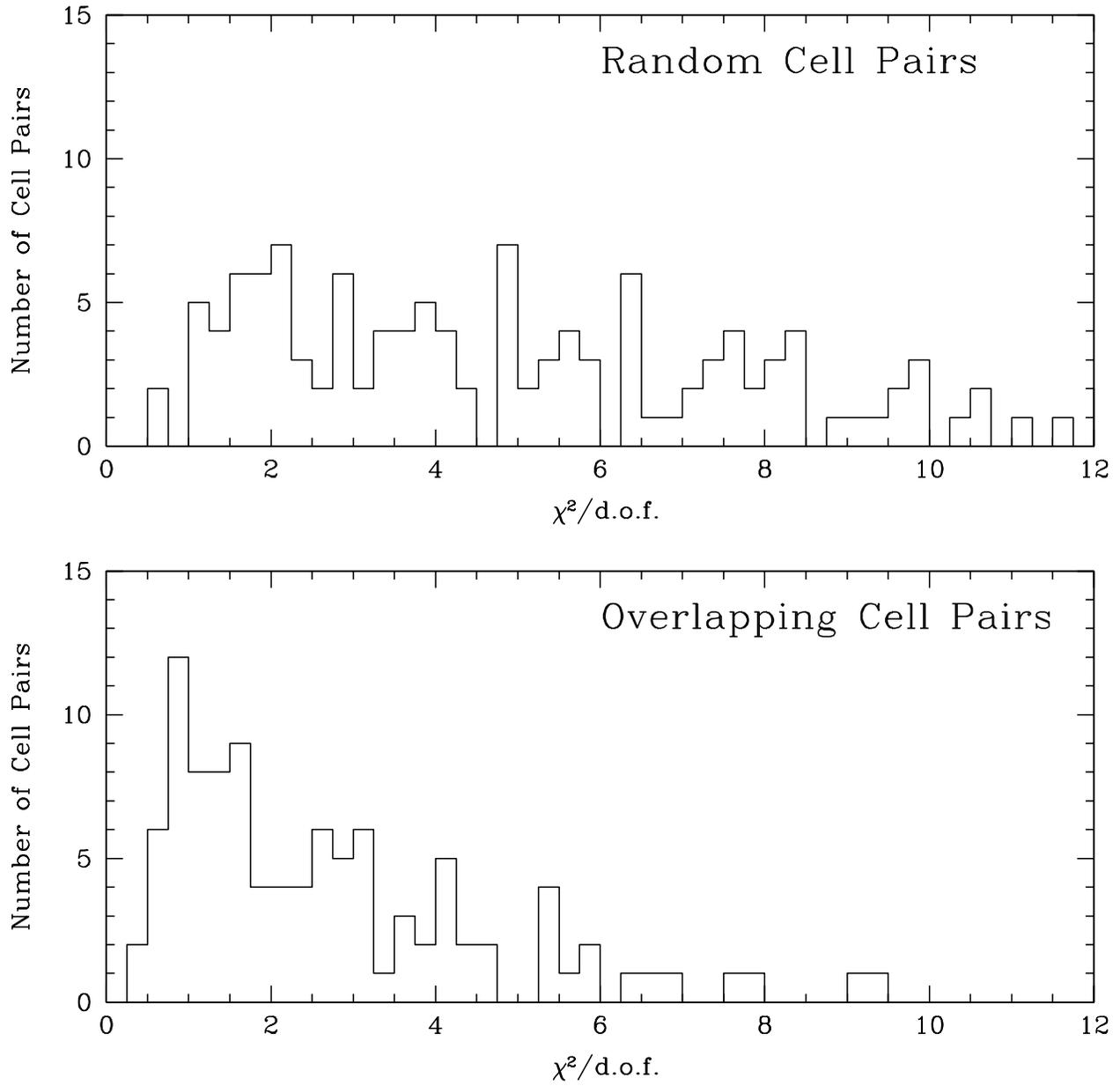,height=7.0in,angle=0}}
\caption{Top: The distribution of $\chi^2$ values measured for random
non-overlapping cell pairs.  Bottom: The distribution of $\chi^2$
values measured for overlapping cell pairs.}
\end{figure}

\clearpage
\topmargin -1.0cm
\begin{figure}
\figurenum{10}
\centerline{\psfig{file=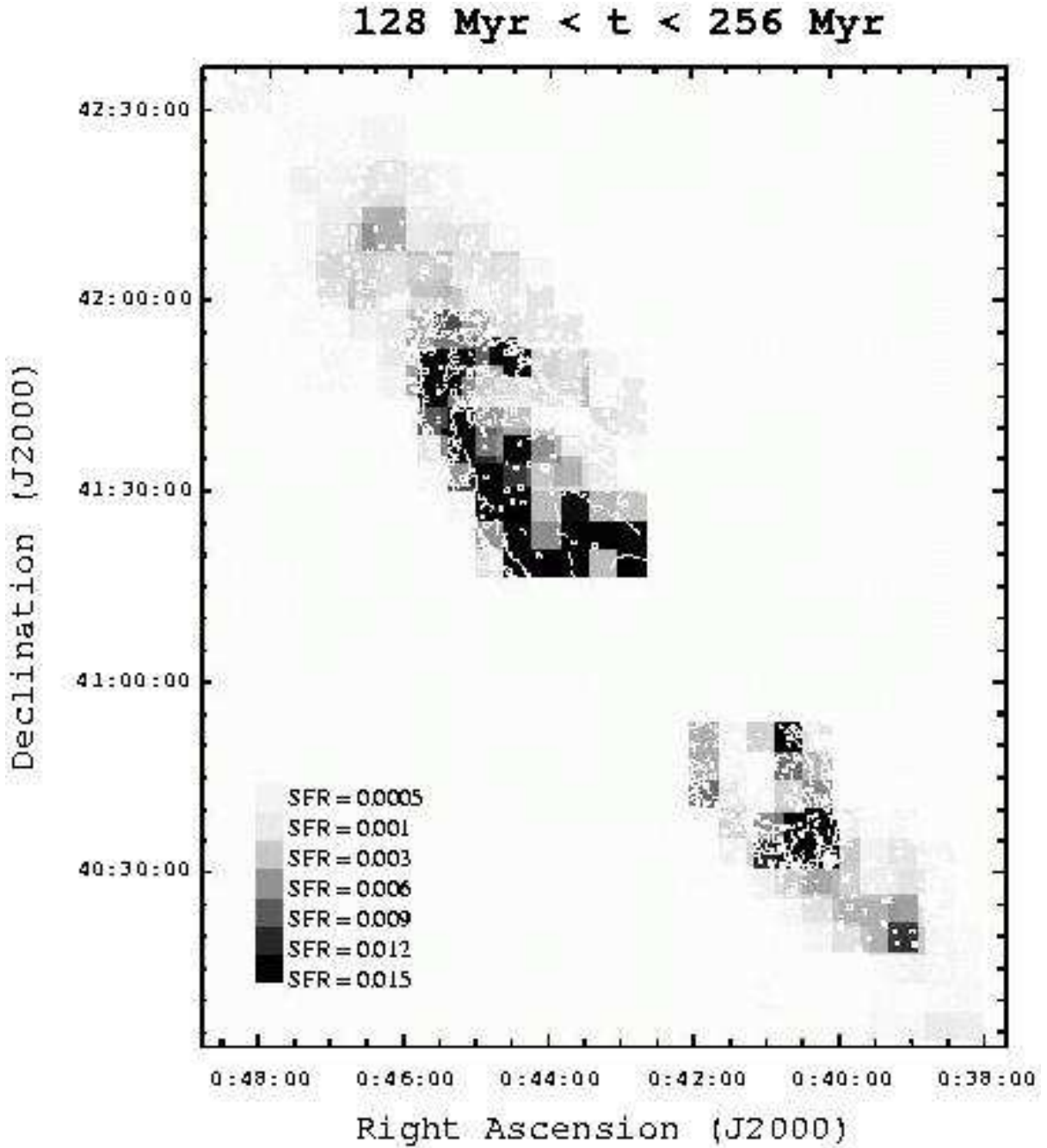,height=7.0in,angle=0}}
\caption{This six-panel figure shows the evolution of star formation
in all of the regions analyzed for the past 256 Myr.  Globally, the
star formation rate over the past 100 Myr appears lower than that of
the previous 100 Myr.  While areas of strong star formation have
stayed quite consistent, there is a suggestion of slow propagation of
star formation from south to north in the southwestern disk and from
southwest to northeast in the northwestern disk: all the same
direction as the rotation of the M31 disk.  The northeast spiral arm
appears to have been more active $\gap$100 Myr ago.  Then it
experienced a star formation lull, and very recently activity has
increased again, suggesting interaction with multiple spiral density
waves.  All of the star formation rates, provided in the greyscale
keys for each map, are in units of 0.054 $M_{\odot} yr^{-1}
arcmin^{-2}$.  These units arise from the size of the cells, which are
each 18.5 arcmin$^{2}$.}
\end{figure}

\begin{figure}
\figurenum{10}
\centerline{\psfig{file=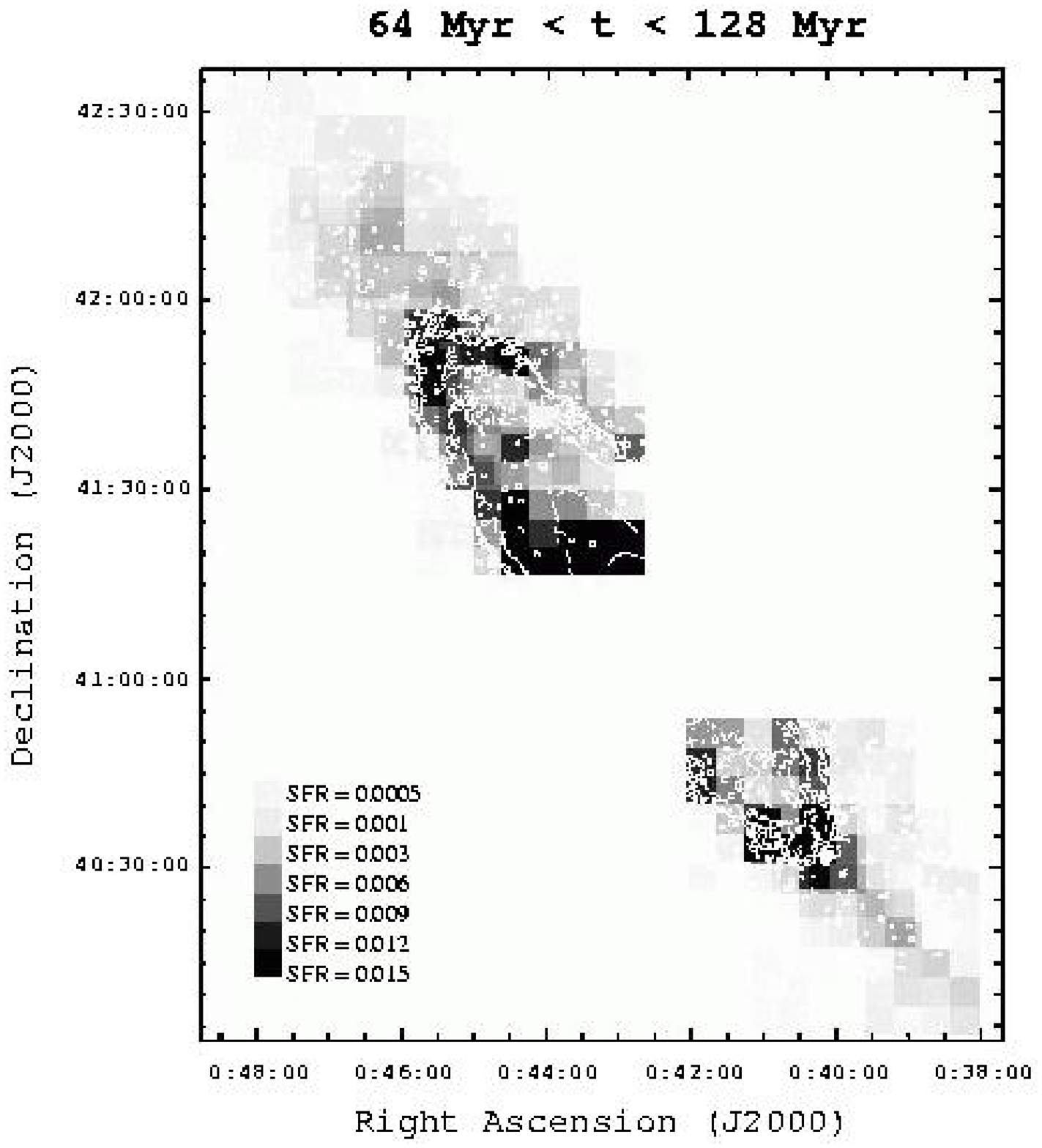,height=7.0in,angle=0}}
\caption{continued}
\end{figure}

\begin{figure}
\figurenum{10}
\centerline{\psfig{file=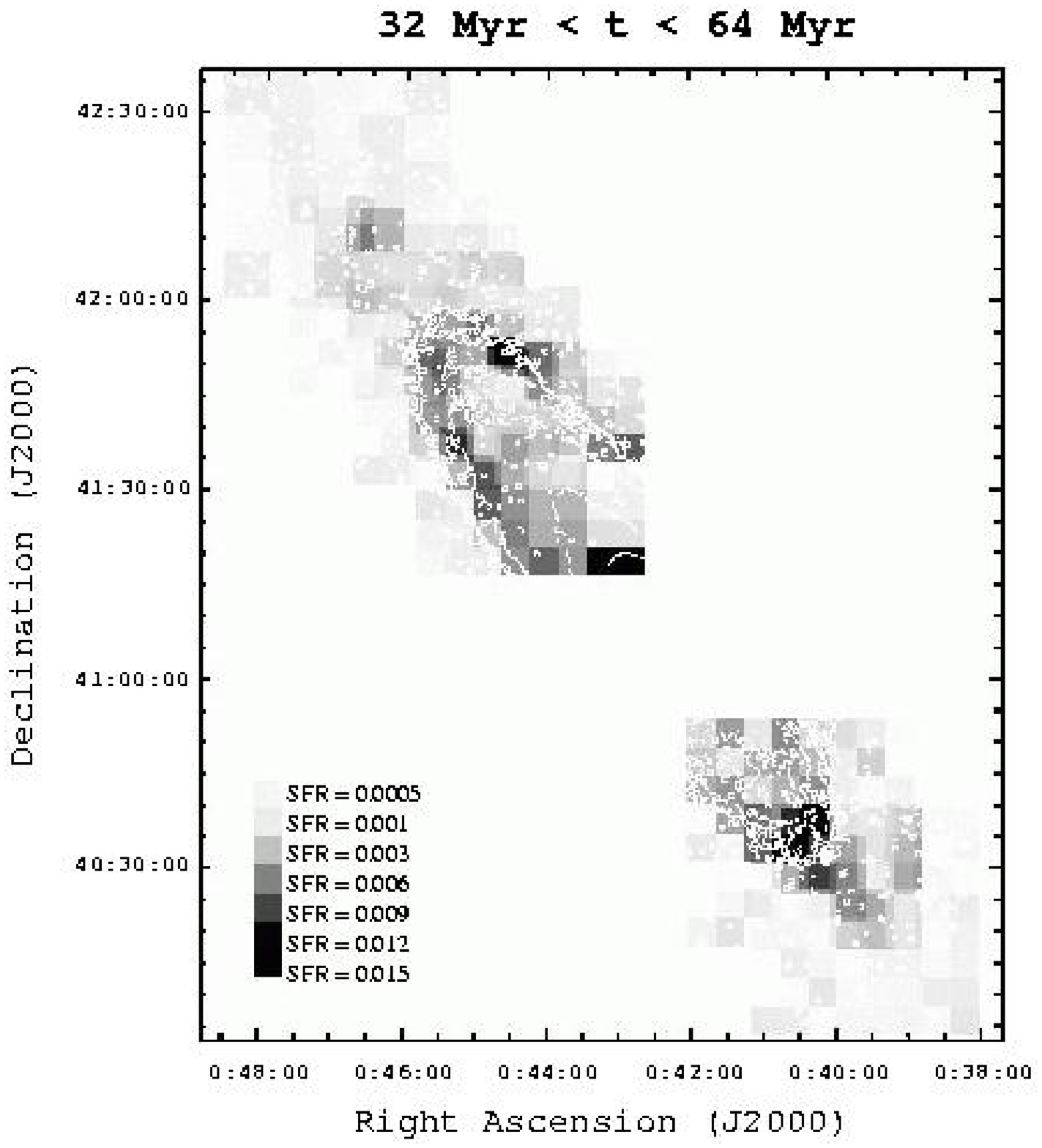,height=7.0in,angle=0}}
\caption{continued}
\end{figure}

\begin{figure}
\figurenum{10}
\centerline{\psfig{file=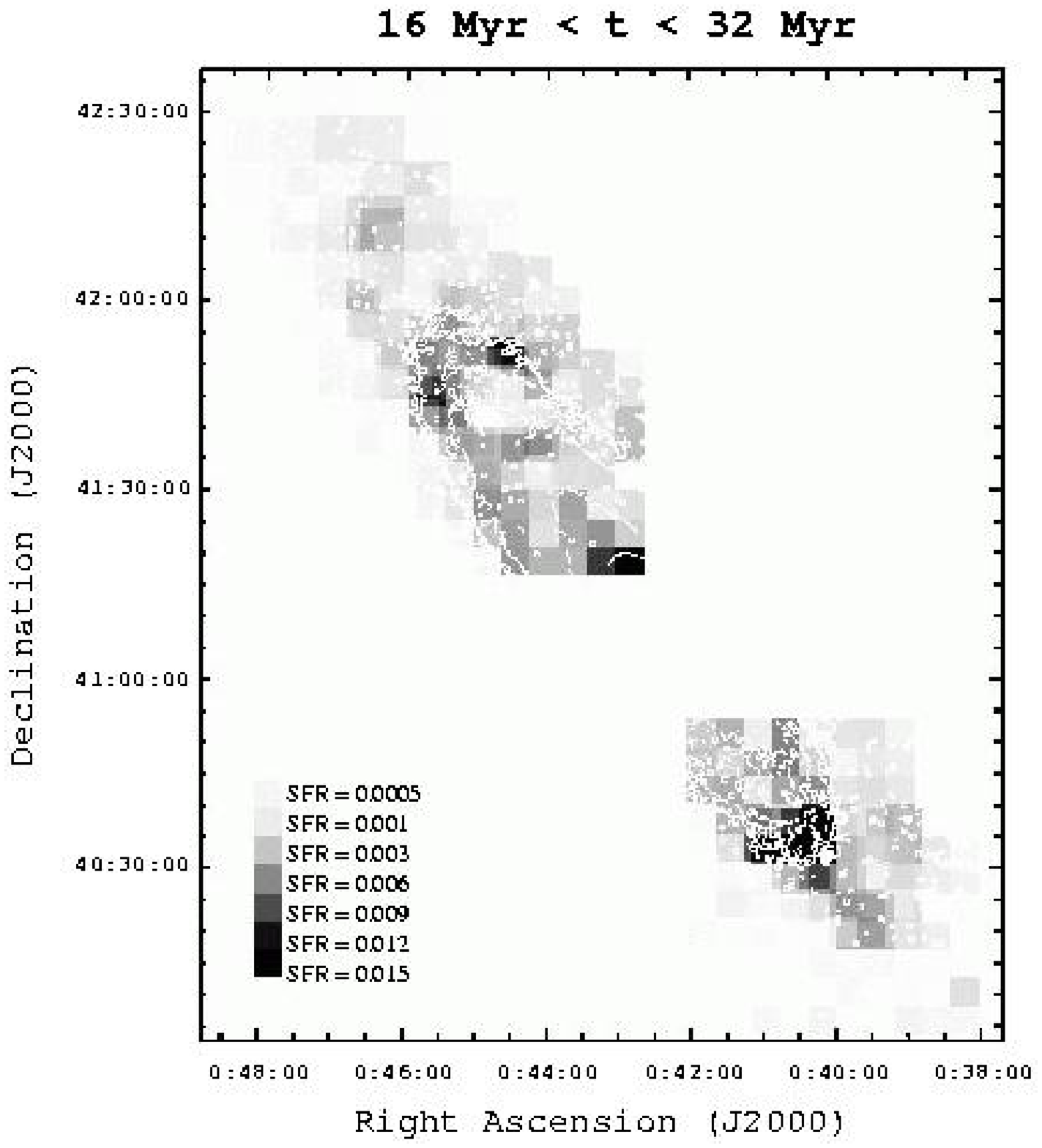,height=7.0in,angle=0}}
\caption{continued}
\end{figure}

\begin{figure}
\figurenum{10}
\centerline{\psfig{file=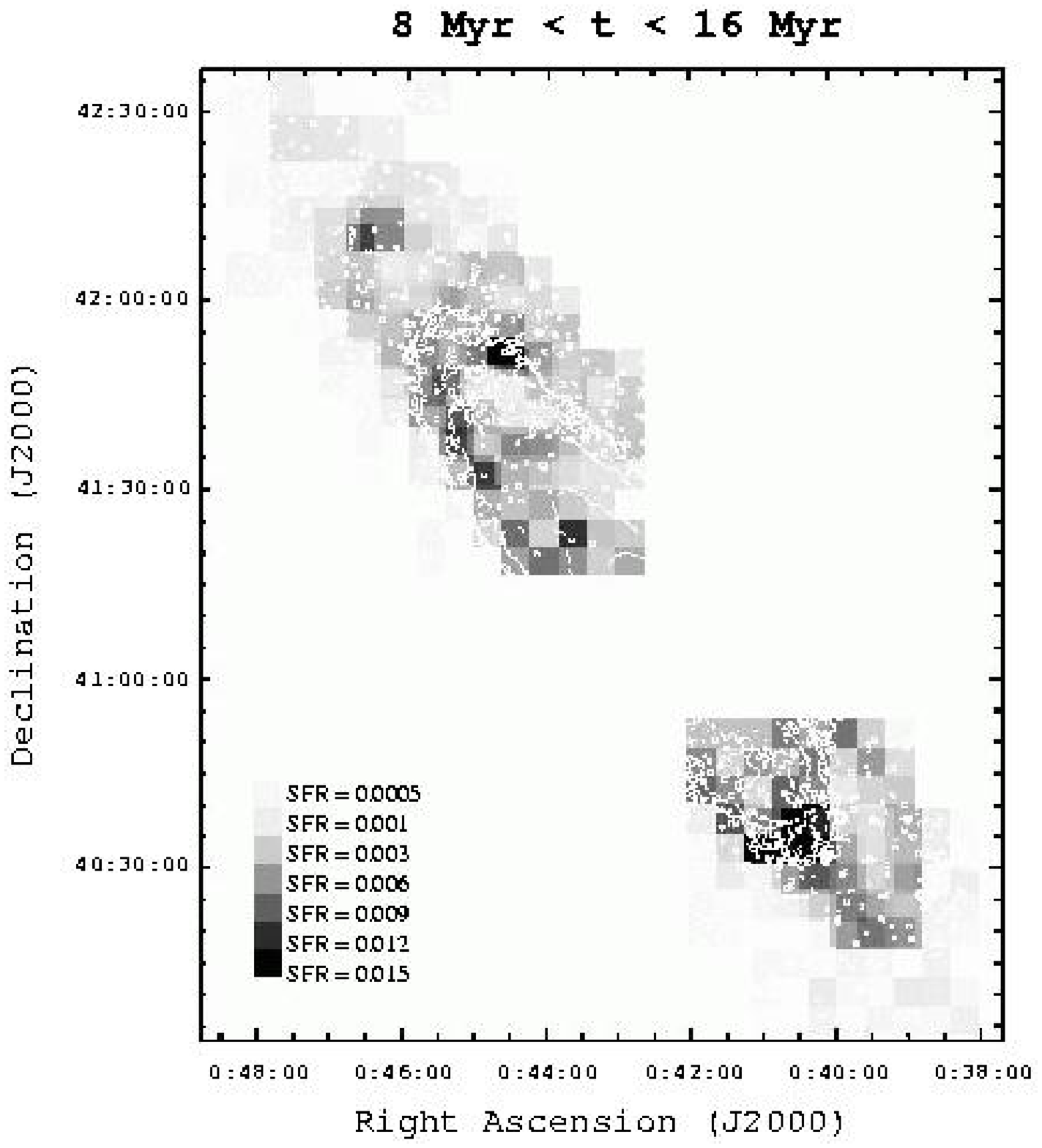,height=7.0in,angle=0}}
\caption{continued}
\end{figure}

\begin{figure}
\figurenum{10}
\centerline{\psfig{file=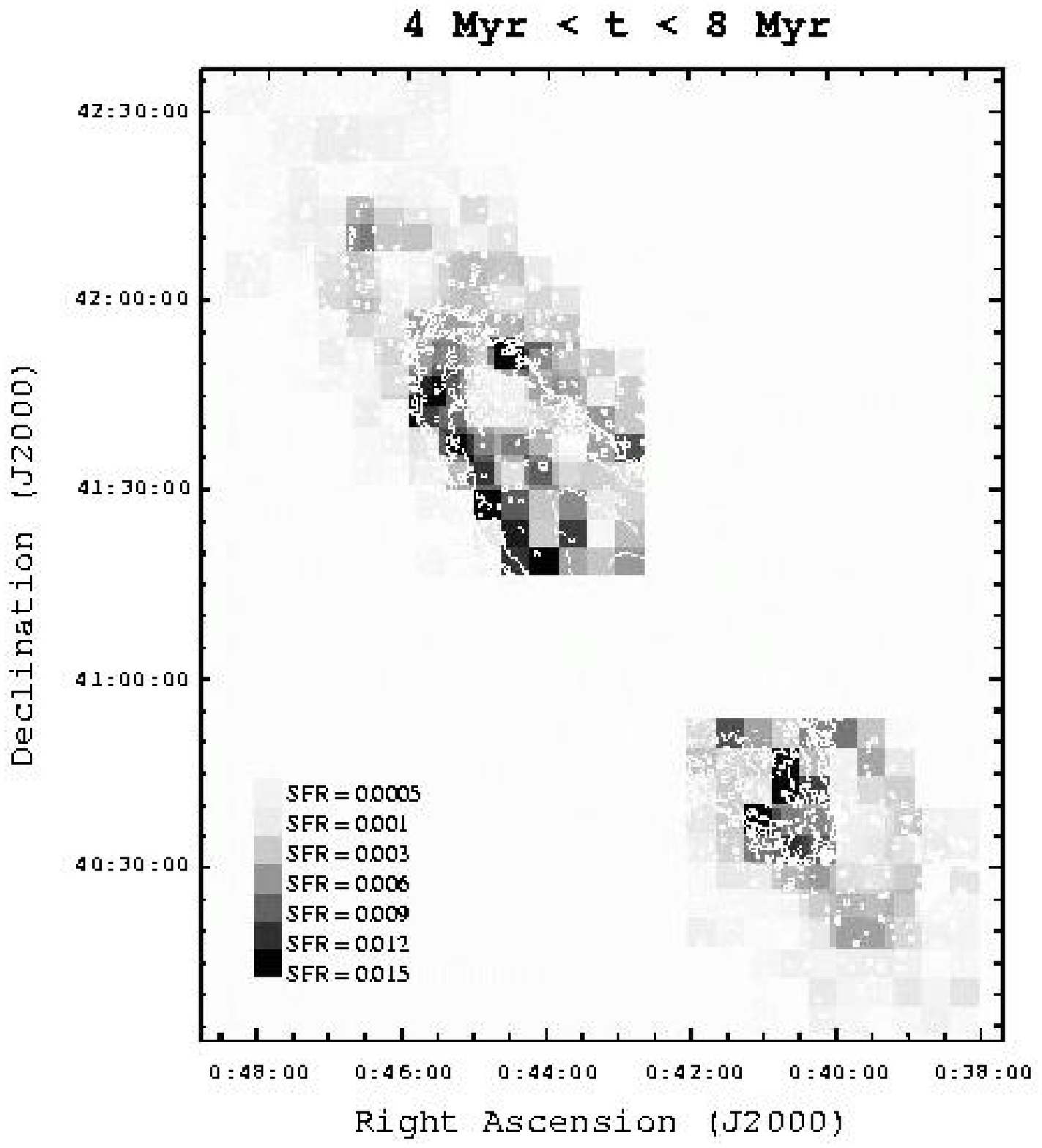,height=7.0in,angle=0}}
\caption{continued}
\end{figure}

\clearpage
\begin{deluxetable}{cccc}
\tablewidth{4.0in}
\tablecaption{$\chi^2$ measurements for comparing SFH results from overlapping cell pairs.}
\tableheadfrac{0.01}
\tablehead{
\colhead{\bf{Cell Pair}} &
\colhead{\bf{R\tablenotemark{a} (arcmin)}} &
\colhead{\bf{$\chi^2$\tablenotemark{b}}} &
\colhead{\bf{$\chi^2$\tablenotemark{c}}}
}
\tablenotetext{a}{Distance between cell centers}
\tablenotetext{b}{All epochs included}
\tablenotetext{c}{One outlier epoch excluded}
\startdata
F1~3,1 and F2~2,4 & 2.03 & 1.25 & 0.87\\
F1~3,1 and F2~2,5 & 2.68 & 0.90 & 0.66\\
F1~3,2 and F2~2,6 & 0.69 & 2.37 & 1.35\\
F1~3,3 and F2~2,7 & 2.89 & 1.36 & 1.36\\
F1~3,3 and F2~2,8 & 1.82 & 1.51 & 0.11\\
F1~4,1 and F3~1,7 & 2.95 & 1.33 & 1.33\\
F1~4,1 and F3~1,8 & 2.51 & 2.08 & 1.30\\
F1~4,2 and F2~3,6 & 2.96 & 0.40 & 0.40\\
F1~5,1 and F2~5,4 & 1.98 & 0.70 & 0.63\\
F1~5,1 and F2~5,5 & 2.65 & 1.54 & 1.26\\
F1~5,1 and F3~3,8 & 1.81 & 0.99 & 0.84\\
F1~5,2 and F2~5,6 & 0.54 & 2.58 & 2.58\\
F1~5,3 and F2~5,7 & 2.85 & 2.11 & 1.76\\
F2~3,1 and F3~1,4 & 1.54 & 0.52 & 0.34\\
F2~3,2 and F3~1,5 & 1.54 & 0.69 & 0.14\\
F2~3,3 and F3~1,6 & 1.54 & 2.88 & 1.92\\
F2~3,4 and F3~1,7 & 1.54 & 0.80 & 0.34\\
F2~3,5 and F3~1,8 & 1.55 & 4.26 & 2.62\\
F2~4,1 and F3~2,4 & 1.54 & 5.33 & 3.79\\
F2~4,2 and F3~2,5 & 1.54 & 1.70 & 1.44\\
F2~4,3 and F3~2,6 & 1.54 & 2.19 & 1.61\\
F2~4,4 and F3~2,7 & 1.54 & 6.25 & 2.23\\
F2~4,5 and F3~2,8 & 1.55 & 9.31 & 2.15\\
F2~5,1 and F3~3,4 & 1.54 & 2.33 & 2.17\\
F2~5,2 and F3~3,5 & 1.54 & 2.92 & 1.14\\
F2~5,2 and F3~3,6 & 2.28 & 2.67 & 1.58\\
F2~5,2 and F4~1,8 & 3.46 & 1.75 & 1.48\\
F2~5,3 and F3~3,6 & 2.92 & 3.63 & 2.37\\
F2~5,3 and F3~3,7 & 2.28 & 6.87 & 4.34\\
F2~5,4 and F3~3,7 & 2.92 & 0.98 & 0.68\\
F2~5,4 and F3~3,8 & 2.28 & 4.13 & 0.65\\
F2~5,5 and F3~3,8 & 2.92 & 9.05 & 2.01\\
F2~6,1 and F3~4,4 & 1.54 & 2.98 & 2.02\\
F2~6,1 and F4~2,7 & 3.46 & 3.47 & 2.29\\
F2~6,2 and F3~4,5 & 1.54 & 3.80 & 2.31\\
\\
\\
F2~6,2 and F4~2,8 & 3.46 & 1.33 & 0.41\\
F2~6,3 and F3~4,6 & 1.54 & 1.51 & 1.17\\
F2~6,4 and F3~4,7 & 1.54 & 1.46 & 1.24\\
F2~6,5 and F3~4,8 & 1.55 & 2.81 & 1.20\\
F2~7,1 and F3~5,4 & 1.54 & 3.15 & 2.11\\
F2~7,2 and F3~5,5 & 1.54 & 3.12 & 0.76\\
F2~7,2 and F4~3,8 & 3.46 & 2.01 & 1.35\\
F2~7,3 and F3~5,6 & 1.54 & 1.03 & 0.44\\
F2~7,4 and F3~5,7 & 1.54 & 4.16 & 1.06\\
F2~7,5 and F3~5,8 & 1.54 & 5.80 & 2.28\\
F2~8,1 and F3~6,4 & 1.54 & 5.36 & 1.23\\
F2~8,2 and F3~6,5 & 1.54 & 1.02 & 0.80\\
F2~8,3 and F3~6,6 & 1.54 & 2.65 & 2.17\\
F2~8,4 and F3~6,7 & 1.54 & 3.20 & 0.60\\
F2~8,5 and F3~6,8 & 1.54 & 4.39 & 2.52\\
F3~3,1 and F4~1,4 & 1.93 & 17.05 & 4.33\\
F3~3,2 and F4~1,5 & 1.94 & 3.04 & 1.39\\
F3~3,3 and F4~1,6 & 1.94 & 5.29 & 4.67\\
F3~3,4 and F4~1,7 & 1.94 & 2.78 & 1.30\\
F3~3,5 and F4~1,8 & 1.94 & 3.42 & 2.14\\
F3~4,1 and F4~2,4 & 1.93 & 1.89 & 1.03\\
F3~4,2 and F4~2,5 & 1.93 & 1.00 & 0.79\\
F3~4,3 and F4~2,6 & 1.94 & 4.03 & 2.97\\
F3~4,4 and F4~2,7 & 1.94 & 1.29 & 1.04\\
F3~4,5 and F4~2,8 & 1.94 & 1.18 & 0.95\\
F3~5,1 and F4~3,4 & 1.93 & 0.39 & 0.33\\
F3~5,2 and F4~3,5 & 1.93 & 16.79 & 0.63\\
F3~5,3 and F4~3,6 & 1.94 & 1.02 & 0.74\\
F3~5,4 and F4~3,7 & 1.94 & 1.09 & 0.65\\
F3~5,5 and F4~3,8 & 1.94 & 7.76 & 2.34\\
F3~6,1 and F4~4,4 & 1.93 & 0.52 & 0.38\\
F3~6,2 and F4~4,5 & 1.93 & 0.72 & 0.46\\
F3~6,3 and F4~4,6 & 1.93 & 1.15 & 0.94\\
F3~6,4 and F4~4,7 & 1.94 & 1.93 & 1.50\\
F3~6,5 and F4~4,8 & 1.94 & 1.48 & 1.33\\
F3~7,1 and F4~5,4 & 1.93 & 1.55 & 1.33\\
F3~7,2 and F4~5,5 & 1.93 & 0.64 & 0.51\\
F3~7,3 and F4~5,6 & 1.93 & 7.56 & 2.38\\
F3~7,4 and F4~5,7 & 1.94 & 5.73 & 3.27\\
F3~7,5 and F4~5,8 & 1.94 & 0.86 & 0.30\\
F3~8,1 and F4~6,4 & 1.93 & 4.05 & 3.14\\
F3~8,2 and F4~6,5 & 1.93 & 3.60 & 1.14\\
F3~8,3 and F4~6,6 & 1.93 & 0.97 & 0.44\\
F3~8,4 and F4~6,7 & 1.94 & 30.65 & 17.11\\
F3~8,5 and F4~6,8 & 1.94 & 2.53 & 0.79\\
F8~3,1 and F9~1,4 & 2.02 & 1.96 & 0.49\\
F8~3,2 and F9~1,5 & 2.02 & 0.84 & 0.66\\
F8~3,3 and F9~1,6 & 2.02 & 3.51 & 2.18\\
F8~3,4 and F9~1,7 & 2.03 & 0.97 & 0.44\\
F8~3,5 and F9~1,8 & 2.03 & 5.32 & 3.53\\
F8~4,1 and F9~2,4 & 2.02 & 1.57 & 1.04\\
F8~4,2 and F9~2,5 & 2.02 & 0.84 & 0.19\\
F8~4,3 and F9~2,6 & 2.02 & 4.22 & 1.97\\
F8~4,4 and F9~2,7 & 2.03 & 1.03 & 0.72\\
F8~4,5 and F9~2,8 & 2.03 & 2.43 & 2.12\\
F8~5,1 and F9~3,4 & 2.02 & 1.46 & 1.01\\
F8~5,2 and F9~3,5 & 2.02 & 2.73 & 2.02\\
F8~5,3 and F9~3,6 & 2.02 & 2.75 & 0.24\\
F8~5,4 and F9~3,7 & 2.03 & 1.31 & 0.95\\
F8~5,5 and F9~3,8 & 2.03 & 1.63 & 1.44\\
F8~6,1 and F9~4,4 & 2.02 & 3.98 & 3.31\\
F8~6,2 and F9~4,5 & 2.02 & 3.21 & 2.34\\
F8~6,3 and F9~4,6 & 2.02 & 4.67 & 2.49\\
F8~6,4 and F9~4,7 & 2.03 & 5.81 & 2.36\\
F8~6,5 and F9~4,8 & 2.03 & 2.11 & 1.73\\
F8~7,1 and F9~5,4 & 2.02 & 4.57 & 1.65\\
F8~7,2 and F9~5,5 & 2.02 & 6.51 & 0.94\\
F8~7,3 and F9~5,6 & 2.02 & 1.99 & 1.32\\
F8~7,4 and F9~5,7 & 2.02 & 1.66 & 0.94\\
F8~7,5 and F9~5,8 & 2.03 & 2.42 & 1.25\\
F8~8,1 and F9~6,4 & 2.02 & 0.79 & 0.64\\
F8~8,2 and F9~6,5 & 2.02 & 0.83 & 0.69\\
F8~8,3 and F9~6,6 & 2.02 & 3.09 & 2.00\\
F8~8,4 and F9~6,7 & 2.02 & 1.09 & 0.68\\
F8~8,5 and F9~6,8 & 2.03 & 1.53 & 0.99\\
\enddata
\end{deluxetable}
\clearpage
\begin{deluxetable}{ccc}
\tablewidth{4.0in}
\tablecaption{Total Star Formation Rate of the Analyzed Area (1.4 deg$^2$) of M31 over 256 Myr}
\tableheadfrac{0.01}
\tablehead{
\colhead{\bf{Epoch}} &
\colhead{\bf{Time (Myr ago)}} &
\colhead{\bf{SFR (M$_{\odot}\ yr^{-1}$)}}
}
\startdata
1 & 256$>$t$>$128 & 1.32$\pm$0.29\\
2 & 128$>$t$>$64 & 1.54$\pm$0.16\\
3 & 64$>$t$>$32 & 0.62$\pm$0.09\\
4 & 32$>$t$>$16 & 0.57$\pm$0.05\\
5 & 16$>$t$>$8 & 0.76$\pm$0.04\\
6 & 8$>$t$>$4 & 0.71$\pm$0.05\\
\hline
\tablevspace{0.05in}
3-6 & 64$>$t$>$4 & 0.63$\pm$0.07\\
\enddata
\end{deluxetable}

\end{document}